\documentclass[journal]{IEEEtran}
\usepackage{amsmath}
\usepackage{cite}
\usepackage{amssymb}
\usepackage{array}
\usepackage{amsfonts}
\usepackage{algorithm}
\usepackage{algpseudocode}
\usepackage{comment}
\usepackage{dsfont}
\usepackage{graphicx}
\usepackage{epsfig}
\usepackage{subfigure}
\usepackage{float}
\usepackage{epstopdf}
\usepackage{psfrag}
\usepackage{xcolor}
\usepackage{url}
\usepackage[colorlinks,linkcolor=black,urlcolor=black,anchorcolor=black,citecolor=black,hyperfootnotes=true]{hyperref}

\newtheorem{corollary}{Corollary}

\newtheorem{theorem}{Theorem}

\newtheorem{remark}{Remark}

\usepackage{mathtools}
\DeclarePairedDelimiter\ceil{\lceil}{\rceil}

\title{{Intelligent Reflecting Surface Based Localization of Mixed Near-Field and Far-Field Targets}
\thanks{
This paper was presented in part at the IEEE International Conference on Wireless Communications and Signal Processing (WCSP) 2024 \cite{Wang_2024_WCSP}.}
\thanks{W. Zhu, Q. Wang, S. Zhang, and L. Liu are with the Department of Electrical and Electronic Engineering, The Hong Kong Polytechnic University, Hong Kong SAR, China (e-mails: \{eee-wf.zhu, shuowen.zhang, liang-eie.liu\}@polyu.edu.hk, qipeng.wang@connect.polyu.hk).}
\thanks{B. Di is with the Department of Electronics, Peking University, Beijing 100871, China (e-mail: diboya@pku.edu.cn).}
\thanks{Yonina C. Eldar is with the Faculty of Mathematics and Computer Science, The Weizmann Institute of Science, Rehovot 7610001, Israel (e-mail: yonina.eldar@weizmann.ac.il).}
}
\author{\IEEEauthorblockN{Weifeng Zhu, Qipeng Wang, Shuowen Zhang, Boya Di, Liang Liu, and Yonina C. Eldar}}

\begin{document}
		\maketitle 


\begin{abstract}
This paper considers an intelligent reflecting surface (IRS)-assisted bi-static localization architecture for the sixth-generation (6G) integrated sensing and communication (ISAC) network. The system consists of a transmit user, a receive base station (BS), an IRS, and multiple targets in either the far-field or near-field region of the IRS. In particular, we focus on the challenging scenario where the line-of-sight (LOS) paths between targets and the BS are blocked, such that the emitted orthogonal frequency division multiplexing (OFDM) signals from the user reach the BS merely via the user-target-IRS-BS path. Based on the signals received by the BS, our goal is to localize the targets by estimating their relative positions to the IRS, instead of to the BS. We show that subspace-based methods, such as the multiple signal classification (MUSIC) algorithm, can be applied onto the BS's received signals to estimate the relative states from the targets to the IRS. To this end, we create a virtual signal via combining user-target-IRS-BS channels over various time slots. By applying MUSIC on such a virtual signal, we are able to detect the far-field targets and the near-field targets, and estimate the angle-of-arrivals (AOAs) and/or ranges from the targets to the IRS. Furthermore, we theoretically verify that the proposed method can perfectly estimate the relative states from the targets to the IRS in the ideal case with infinite coherence blocks. Numerical results verify the effectiveness of our proposed IRS-assisted localization scheme. Our paper demonstrates the potential of employing passive anchors, i.e., IRSs, to improve the sensing coverage of the active anchors, i.e., BSs.

\end{abstract} 
\begin{IEEEkeywords}
Integrated sensing and communication (ISAC), intelligent reflecting surface (IRS), orthogonal frequency division multiplexing (OFDM), the sixth-generation (6G) cellular network, mixed near-field and far-field targets, multiple signal classification (MUSIC).
\end{IEEEkeywords}

\section{Introduction}


Integrated sensing and communication (ISAC) is one of the primary use cases of the future sixth-generation (6G) cellular systems \cite{itu} and has attracted tremendous attention in the literature \cite{isac_survey1,isac_survey2,Zhang_2021_JSTSP,Liu_2022_CST}. It is expected that the future 6G network will provide not only high-quality communication services, but also new sensing functionalities such as high-accuracy localization, large-scale tracking, and high-resolution imaging. However, how to effectively integrate the above sensing functionalities into a communication-oriented network remains an open problem. 

Among various sensing services in 6G ISAC systems, localization is considered to be one of the most crucial functionalities. It has wide-ranging applications and also serves as the foundation to other sensing tasks such as tracking and imaging. In 6G ISAC systems, base stations (BSs) are expected to be the primary anchors to perform localization, because they are deployed at known positions and can emit and process radio signals for estimating the range and angle-of-arrival (AOA) information of targets \cite{Zhao_2020_TSP,Shen_2012_TWC}. 
However, in many scenarios, line-of-sight (LOS) paths between BSs and targets are blocked by a variety of obstacles, challenging existing BS-centric localization algorithms \cite{dvc}. 

\begin{figure}[t]
	\centering
    \includegraphics[width=.45\textwidth]{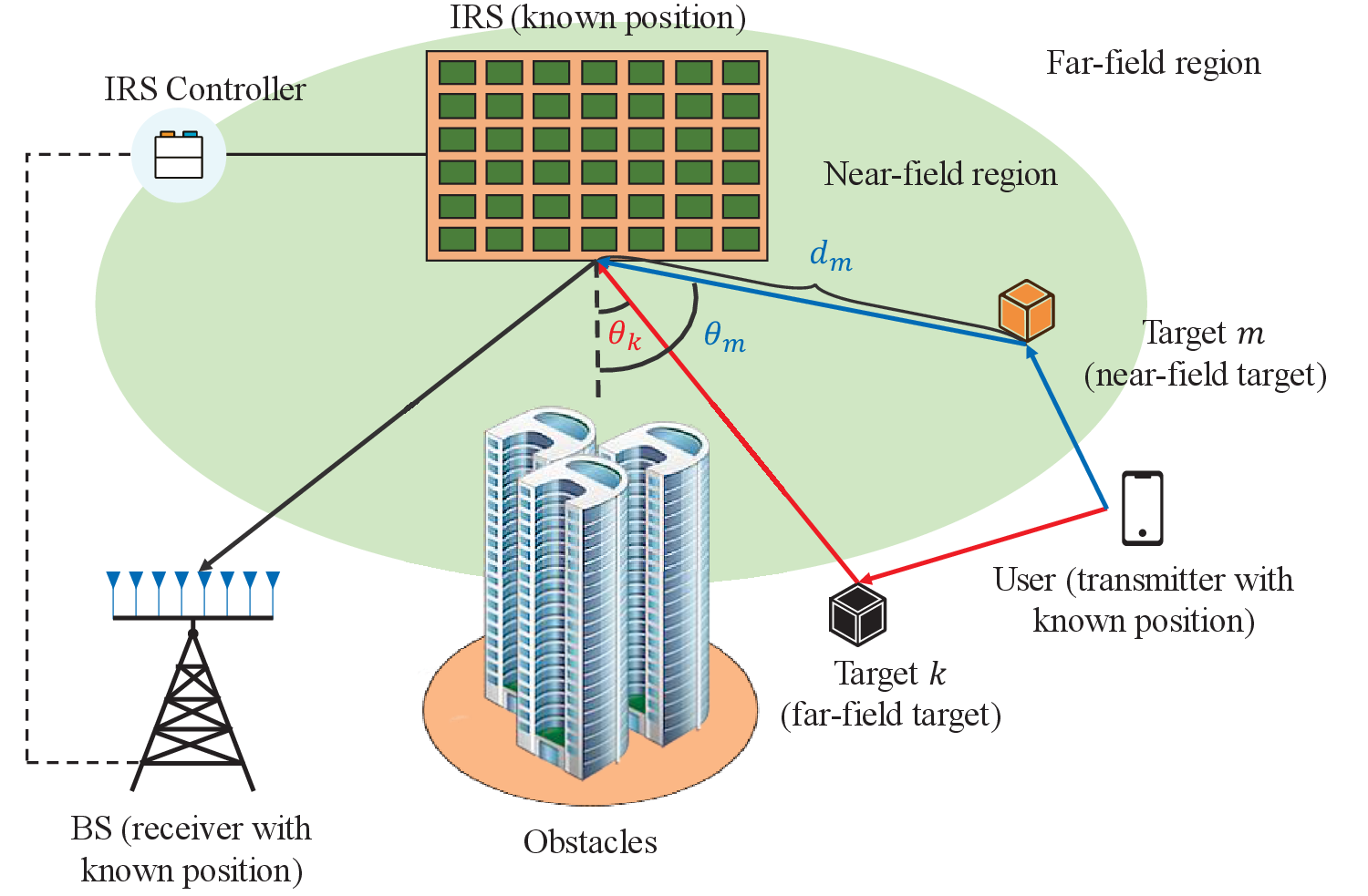}
    \vspace{-0.3cm}
    \caption{System model for IRS-assisted bi-static localization in a 6G ISAC network, where LOS paths between the targets and the BS are blocked, and the BS's received signals are merely over the user-target-IRS-BS path. Therein, targets are in both of the near-field and far-field regions of the IRS. We utilize the IRS as a passive anchor for localization. 
    }\label{fig:system model}
    \vspace{-0.5cm}
\end{figure}

In this paper, we consider the possibility of employing an intelligent reflecting surface (IRS) to localize the targets that are in the non-line-of-sight (NLOS) region of the BS. Specifically, as shown in Fig. \ref{fig:system model}, we study a bi-static localization system in a 6G network, which consists of a single-antenna user as the transmitter, a multi-antenna BS as the receiver, an IRS to assist the BS, and multiple passive targets to be localized. It is assumed that the LOS paths between the targets and the BS are blocked, while the IRS is deployed at a proper site with LOS paths to the targets and the BS. Moreover, because of the large size of the IRS, some targets are in the near-field region of the IRS, while others are in the far-field region of the IRS \cite{Zhang_2023_CM,near_field_survey}. Under this system, the user emits orthogonal frequency division multiplexing (OFDM) signals in the uplink \cite{3gpp_2024_nr}, while the BS receives the echo signals from the targets merely over the user-target-IRS-BS paths. Our goal is to exploit the BS’s received signals to detect the near-field and far-field targets and localize them based on their range and AOA information in reference to the IRS. This is feasible because the target-IRS-BS channels are functions of these AOAs and distances. Since the reference point to localize the targets is the IRS, but the AOAs and distances from the targets to the IRS are estimated by the BS, the IRS plays the role of a \emph{passive anchor} \cite{Liu_2024_CM} in our considered 6G localization system. 

In a localization system without LOS paths between the targets and the BS as shown in Fig. \ref{fig:system model}, the main challenge for employing the IRS as a passive anchor lies in how to estimate the AOAs and distances from the targets to the IRS based on the signals received by the BS, instead of those received by the IRS (since the passive IRS cannot perform signal processing). A breakthrough was made in our previous works \cite{Wang_2022_GC, Wang_2024_TWC}, which verifies the feasibility to estimate the distances between the passive targets and the IRS for the first time in the literature. Specifically, the BS can estimate the overall propagation delay (distance) from a target to the IRS to the BS. Because the distance between the BS and the IRS is known, the distance between a target and the IRS can be then obtained by subtracting the IRS-BS distance from the target-IRS-BS overall distance. As a result, the remaining scientific problem for IRS-assisted localization as shown in Fig. \ref{fig:system model} is how to estimate the AOAs from the targets to the IRS based on the BS's received signals.


Recently, several early works \cite{Zhang_2024_TWC,Han_2022_JSTSP,Rahal_2024_JSTSP,Teng_2023_JSAC} have targeted at the AOA estimation problem in IRS-assisted localization systems similar to Fig. \ref{fig:system model}. Specifically, \cite{Zhang_2024_TWC,Han_2022_JSTSP,Rahal_2024_JSTSP} focused on a special case with one target in the system. Under the far-field model, \cite{Zhang_2024_TWC} proposed a deep learning (DL) based algorithm for adaptive AOA estimation. Moreover, under the near-field model, maximum likelihood (ML) based methods are developed in \cite{Han_2022_JSTSP,Rahal_2024_JSTSP}. As to AOA estimation with multiple targets, \cite{Teng_2023_JSAC} proposed a message passing-based algorithm that can estimate the AOAs from far-field targets to the IRS. However, there are some key limitations of the above works \cite{Zhang_2024_TWC,Han_2022_JSTSP,Rahal_2024_JSTSP,Teng_2023_JSAC} for AOA estimation in IRS-assisted localization systems.

1. For the DL-based method proposed in \cite{Zhang_2024_TWC}, the AOA estimation performance cannot be guaranteed in theory.

2. For the ML-based single AOA estimation algorithms proposed \cite{Han_2022_JSTSP,Rahal_2024_JSTSP}, their extension to the multi-target case is of extremely high complexity, because a multi-dimension exhaustive search should be done simultaneously to solve the ML problem.

3. For the message passing-based algorithm proposed in \cite{Teng_2023_JSAC}, it does not work when near-field and far-field targets both exist in the system.


In the literature of AOA estimation, the subspace-based method, such as the multiple signal classification (MUSIC) algorithm \cite{Schmidt_1986_TAP}, is the most widely used method, thanks to its superior performance and low complexity. 
It estimates the AOAs of multiple targets based on the orthogonality between the signal subspace and the noise subspace in the covariance matrix of received signals. 
If MUSIC can be applied in our considered system shown in Fig. \ref{fig:system model}, it is appealing to tackle all of the limitations of methods proposed in \cite{Zhang_2024_TWC,Han_2022_JSTSP,Rahal_2024_JSTSP,Teng_2023_JSAC}. First, different from the model-free DL method \cite{Zhang_2024_TWC}, tremendous works have theoretically and numerically verify the performance of MUSIC \cite{Swin_1992_TSP,Porat_1988_TASSP}. 
Second, compared with ML-based algorithms in \cite{Han_2022_JSTSP,Rahal_2024_JSTSP} that require a multi-dimension exhaustive search in a multi-target setting, the MUSIC algorithm only requires a one-dimension exhaustive search on the spectrum to estimate all the AOAs. 
Last, different from the message passing-based method [20], MUSIC can be applied in the mixed near-field and far-field localization network \cite{Liang_2010_TSP,Jiang_2013_SJ}. 

Motivated by the above, this paper aims to investigate how to perform the MUSIC algorithm to detect the near-field targets and the far-field targets and estimate the AOAs and/or distances from the targets to the IRS in our considered localization system as shown in Fig. \ref{fig:system model}. However, it is non-trivial to achieve the above goal. Specifically, if we obtain the sample covariance matrix of the signals received by all BS antennas and analyze the so-called ``spatial spectrum" corresponding to this covariance matrix, we can just estimate the AOA from the IRS to the BS. This is because the incident signals of the BS are from the IRS as shown in Fig. \ref{fig:system model}. How to modify the MUSIC algorithm such that it can estimate the AOAs and/or distances from the targets to the IRS is a challenge. Moreover, the MUSIC algorithm is mainly applied in narrowband systems. However, current cellular network is a broadband system. How to apply the MUSIC algorithm in an OFDM-based broadband localization system is a challenge.

In this paper, we manage to tackle all the above challenges in our considered localization system shown in Fig. \ref{fig:system model}. The contributions of this paper are listed as below. 

1) We devise a novel three-phase localization protocol tailored for the considered system. In Phase I, the channel impulse response (CIR) of our considered OFDM system, containing the strength and delay information of each channel tap, is estimated. In Phase II, we first estimate the range information of user-target-IRS-BS paths based on the delay information of non-zero channel taps. Subsequently, because the user-target-IRS-BS channel with a delay tap can be viewed as a narrowband channel, we can apply the subspace-based method to detect the far-field and near-field targets and estimate their AOAs (far-field targets) or AOAs plus ranges (near-field targets) to the IRS. In Phase III, all targets are localized based on the range and AOA estimations obtained in Phase II.

2) The main challenge of our proposed three-phase protocol lies in Phase II - based on the channels associated with the user-target-IRS-BS paths estimated in Phase I, how to utilize the MUSIC algorithm to detect the far-field and near-field targets and estimate their AOAs and/or ranges to the IRS. As mentioned in the above, if we directly apply the MUSIC algorithm to the estimated user-target-IRS-BS channel vectors, the AOA from the IRS to the BS will be estimated, which is not of our interests. We propose a novel approach to tackle this challenge. Specifically, we find a way to design the IRS reflection coefficients over different time slots of the same coherence block such that after combining the user-target-IRS-BS channels over several time slots of a coherence block into a virtual channel vector, this new channel vector can be surprisingly expressed as a function of the virtual steering vectors of the IRS towards the targets. Consequently, we can apply the MUSIC algorithm onto this properly designed virtual vector to accomplish the goals of Phase II under our proposed localization scheme. In particular, the proposed solution can be considered as a generalization of the method proposed in \cite{music_irs}, which considers narrowband systems with far-field targets.

3) Other than signal processing design, we also provide the theoretical performance analysis for the MUSIC algorithm proposed in Phase II of our scheme. In the ideal case with infinite coherence blocks, we rigorously show that the AOA information of far-field targets and the AOA and range information of near-field targets can be perfectly estimated under the subspace-based method. To our best knowledge, this is the first theoretical result to justify the effectiveness of employing IRS as the passive anchor to localize the targets. Moreover, numerical results also show that our proposed subspace-based localization scheme outperforms other non-subspace-based counterparts.

The rest of the paper is organized as follows. Section II introduces the system model of the IRS-assisted sensing system. In Section III, the three-phase localization protocol is presented. Then the specific target detection scheme in Phase II is proposed to extract the AOA and range information of the near-field and far-field targets in Section IV. Finally, Section V evaluates the performance of the proposed localization methods and Section VII concludes this work.

{\it Notations}: In this paper, vectors and matrices are denoted by boldface lower-case letters and boldface upper-case letters, respectively. For a complex vector $\boldsymbol{x}$, $||\boldsymbol{x}||_q$ and $x_n$ denote the $l_q$-norm and the $n$-th element, respectively. For an $M \times N$ matrix $\boldsymbol{X}$, $\boldsymbol{X}^T$ and $\boldsymbol{X}^{H}$, $\mathcal{N}(\boldsymbol{X})$ denote its transpose, conjugate transpose, and null space, respectively. For a complex number $x \in \mathbb{C}$, its phase is denoted as $\angle x$. Further, $\mathbb{E}_{\boldsymbol{a}}[\cdot]$ denotes the expectation operation over random vector $\boldsymbol{a}$.

\section{System Model}\label{sec:system_model}

We consider the localization task in an IRS-assisted 6G cellular system as illustrated in Fig. \ref{fig:system model}, which consists of one single-antenna user as the transmit anchor, one BS equipped with $M_{\text{B}} \ge 1$ antennas as the receive anchor, and $K$ passive targets to be localized. 
The 2D coordinates of the user, the BS, and the $k$-th target are denoted as $(x_{\text{U}},y_{\text{U}})$, $(x_{\text{B}},y_{\text{B}})$, and $(x_k,y_k)$, $\forall k$, respectively. We focus on a challenging scenario where the LOS paths between the targets and the BS are blocked by the obstacles such that the targets cannot be localized via the user-target-BS path as in the conventional bi-static radar system. To tackle this challenge, one IRS equipped with $M_\text{I}$ reflecting elements is deployed at a proper location, denoted by $(x_{\text{I}},y_{\text{I}})$, with LOS paths to the targets and the BS, as shown in Fig. \ref{fig:system model}. Under the above setup, the user emits OFDM signals in the uplink \cite{3gpp_2024_nr}, which will be reflected to the BS via the user-target-IRS-BS path. The BS can perform ISAC to its received signals via decoding user messages and localizing the targets simultaneously. Since IRS-assisted communication has been widely studied in the literature, this paper mainly focuses on localizing the targets based on the signals received at the BS. 

\subsection{Signal Propagation Model}

In this paper, we consider the transmission of OFDM symbols over $V$ consecutive coherence blocks, where each coherence block consists of $Q$ OFDM symbols. Moreover, we assume that there are $N$ OFDM sub-carriers and the sub-carrier spacing is $\Delta f$ Hz such that the bandwidth is $B = N\Delta f$ Hz. In each coherence block $t$, let $s_{n,t}^{(q)}$ denote the pilot signal of the user at the $n$-th sub-carrier in each OFDM symbol $q$, $\forall t, n, q$. Then, the frequency-domain pilot signal generated by the user over all the $N$ sub-carriers of the $q$-th symbol in coherence block $t$ is denoted by $\boldsymbol{s}_t^{(q)}=[s_{1,t}^{(q)},\cdots,s_{N,t}^{(q)}]^T$. Assume that the user transmits with identical power at all sub-carriers, which is denoted by $p$. Then, in each coherence block $t$, the time-domain OFDM signal generated by the user in the $q$-th OFDM symbol duration is denoted by
\begin{align}\label{eq:tx signal}
    \boldsymbol{\chi}_t^{(q)}= [\chi_{1,t}^{(q)},\cdots,\chi_{N,t}^{(q)}]^T = \boldsymbol{W}^H\sqrt{p}\boldsymbol{s}_t^{(q)},\quad \forall q,t, 
\end{align}
where $\chi_{n,t}^{(q)}$ denotes the $n$-th temporal-domain sample from the user at the $q$-th OFDM symbol duration in coherence block $t$, and $\boldsymbol{W}$ is the $N\times N$ discrete Fourier transform (DFT) matrix. At the beginning of each OFDM symbol $q$, a cyclic prefix (CP) consisting of $J$ OFDM samples is inserted to eliminate the inter-symbol interference. Therefore, in coherence block $t$, the overall time-domain pilot signal transmitted by the user in the $q$-th OFDM symbol duration is denoted by
\begin{align}\label{eq:tx signal with cp}
    \boldsymbol{\bar{\chi}}_t^{(q)}=[\chi_{N-J+1,t}^{(q)}, \ldots, \chi_{N,t}^{(q)}, \chi_{1,t}^{(q)}, \ldots, \chi_{N,t}^{(q)}]^T, \quad\forall q,t.
\end{align}

Define $\boldsymbol{\bar{G}}\in\mathbb{C}^{M_{\text{B}}\times M_{\text{I}}}$ as the channel from the IRS to the BS, and $\boldsymbol{r}_{k,t} \in \mathbb{C}^{M_{\text{I}} \times 1}$ as the channel of the user-target $k$-IRS path in coherence block $t$, $\forall k,t$. Moreover, define $\phi_{m_\text{I},t}^{(q)}$ as the reflecting coefficient of the $m_\text{I}$-th IRS element with $|\phi_{m_\text{I},t}^{(q)}| = 1$ during the $q$-th OFDM symbol duration in coherence block $t$, and $\boldsymbol{\phi}_t^{(q)}=[\phi_{1,t}^{(q)},\cdots,\phi_{M_\text{I},t}^{(q)}]^T \in \mathbb{C}^{M_\text{I}\times 1}$. Therefore, the effective channel of the user-target $k$-IRS-BS path at the $q$-th OFDM symbol transmission in coherence block $t$ can be defined as
\begin{align}\label{eq:cascaded channel}
    \boldsymbol{h}_{k,t}^{(q)} = \boldsymbol{\bar{G}}\text{diag}(\boldsymbol{\phi}_t^{(q)})\boldsymbol{r}_{k,t}\in\mathbb{C}^{M_{\text{B}} \times 1}, \quad \forall q,t,k.
\end{align}
Then, at the $n$-th sample period of the $q$-th OFDM symbol duration in coherence block $t$, the signal received by the BS is given as\footnote{In the case that the LOS path between the user and the IRS is not blocked, the signals can also be propagated over the user-IRS-BS path. Since the user, the IRS, and the BS are deployed at fixed and known locations, the LOS user-IRS channel and IRS-BS channel are known \cite{assup_known_channel1,assup_known_channel2}. As a result, the signals via the user-IRS-BS path can be canceled from the received signal such that (\ref{eq:time domain received signal}) can be interpreted as the processed received signal that is useful for localizing the targets.}
\vspace{-10pt}
\begin{align}\label{eq:time domain received signal}
    \boldsymbol{\upsilon}_{n,t}^{(q)} = \sum_{k=1}^{K}\boldsymbol{h}_{k,t}^{(q)}\bar{\chi}_{n-l_k,t}^{(q)} + \boldsymbol{z}_{n,t}^{(q)},\quad\forall n,q,t,
\end{align}where $l_k$ denotes the propagation delay (in terms of OFDM samples) of the user-target $k$-IRS-BS path, and $\boldsymbol{z}_{n,t}^{(q)}\sim \mathcal{CN}(0,\sigma^2\boldsymbol{I})\in\mathbb{C}^{M_{\text{B}} \times 1}$ is the Gaussian noise of the BS at the $n$-th sample period of the $q$-th OFDM symbol duration in coherence block $t$. In an $L$-tap multi-path environment where $L$ is the maximum number of resolvable paths, define 
\begin{align}\label{eq:range cluster l}
    \Omega_l = \{k| l_k = l\},~l=1,\ldots,L,
\end{align}
as the set of targets in range cluster $l$, i.e., for all the targets in $\Omega_l$, their echo signals will be propagated to the BS simultaneously with a delay of $l$ OFDM samples. Moreover, define $K_l=|\Omega_l|$ as the number of targets in range cluster $l$. $\forall l$. At the $q$-th OFDM symbol duration in coherence block $t$, define the effective channel from the user to all the targets in range cluster $l$ to the IRS to the BS as
\begin{align}\label{eq:multipath channel l tap}
\boldsymbol{\bar{h}}_{l,t}^{(q)} = \left\{\begin{matrix}
\sum_{k\in\Omega_l}\boldsymbol{h}_{k,t}^{(q)}, & \text{if}~\Omega_l \neq \emptyset, \\
\boldsymbol{0}, & \text{otherwise},
\end{matrix}\right. \quad\forall q,l,t.
\end{align}Moreover, define
\vspace{-5pt}
\begin{align}\label{eq:multipath channel}
    \boldsymbol{\bar{H}}_t^{(q)} = [\boldsymbol{\bar{h}}_{1,t}^{(q)},\cdots,\boldsymbol{\bar{h}}_{L,t}^{(q)}]^T\in \mathbb{C}^{L\times M_{\text{B}}},\quad \forall q,t,
\end{align}
as the $L$-tap multi-path channel from the user to the BS at the $q$-th OFDM symbol duration in coherence block $t$. Then, the time-domain received signal given in \eqref{eq:time domain received signal} can be expressed as
\vspace{-5pt}
\begin{align}
   \boldsymbol{\upsilon}_{n,t}^{(q)} = \sum_{l=1}^{L}\boldsymbol{\bar{h}}_{l,t}^{(q)}\bar{\chi}_{n-l,t}^{(q)} + \boldsymbol{z}_{n,t}^{(q)},\quad\forall n,q,t.
\end{align}After removing CP and performing the DFT operation to the above time-domain signal, the frequency-domain received signal over all the $N$ sub-carriers of the $q$-th OFDM symbol in coherence block $t$ is given as
\vspace{-5pt}
\begin{align}\label{eq:received signal}
    \boldsymbol{\bar{\Upsilon}}_t^{(q)} & = [\boldsymbol{\bar{\upsilon}}_{1,t}^{(q)},\cdots,\boldsymbol{\bar{\upsilon}}_{N,t}^{(q)}]^T \notag \\ &=  \sqrt{p}~ \text{diag}({\boldsymbol{s}}_t^{(q)})\boldsymbol{E}\boldsymbol{\bar{H}}_t^{(q)} + \boldsymbol{\bar{Z}}_t^{(q)},\quad \forall q, 
\end{align}where $\boldsymbol{E}\in\mathbb{C}^{N\times L}$ with the element on the $n$-th row and $l$-th column denoted by $E_{n,l} = e^{-j\frac{2\pi(n-1)(l-1)}{N}}$, and $\boldsymbol{\bar{Z}}_t^{(q)}=[\boldsymbol{z}_{1,t}^{(q)},\cdots,\boldsymbol{z}_{N,t}^{(q)}]^T\in\mathbb{C}^{N\times M_{\text{B}}}$.

\subsection{Channel Model}\label{subsec:channel model}

It is worth noting that the range and AOA information of all the targets are embedded in the multi-path channels $\boldsymbol{\bar{h}}_{l,t}^{(q)}$'s. In the following, we introduce the channel model that reveals the relation between the target locations and the channels. 

Specifically, because the IRS is physically large, its near-field region is in general large such that some targets are in the near-field region of the IRS, and some targets are in the far-field region of the IRS. Therefore, in this paper, we adopt the general near-field steering vector model at the IRS side. In particular, towards a point of distance $d$ and angle $\eta$, let $\boldsymbol{a}_{\text{I}}(d,\eta) = [a_{\text{I},{1}}(d,\eta),\cdots,a_{\text{I},M_{\text{I}}}(d,\eta)]^T$ denote the near-field steering vector of the IRS. For example, if the IRS elements are deployed under the uniform linear array (ULA) model with element spacing $d^{\text{I}}$ meters, the $m_{\text{I}}$-th response in the steering vector can be characterized as \cite{near_field_survey}
\vspace{-5pt}
\begin{align}\label{eq:a_m}
a_{\text{I},m_{\text{I}}}(d,\eta) &= e^{-j\frac{2\pi}{\lambda}(d-\sqrt{d^2+(m_{\text{I}}d^{\text{I}})^2-2m_{\text{I}}d d^{\text{I}}\cos\eta})}, \notag \\
&\approx e^{-j\frac{2\pi}{\lambda} (m_{\text{I}} d^{\text{I}}\cos\eta + \frac{(m_{\text{I}}d^{\text{I}}\sin\eta)^2}{2d})}, ~\forall m_I,
\end{align}where $\lambda$ in meter is the wavelength. Note that mathematically, the far-field steering vector is a special case of the near-field steering vector when $d$ goes to infinity. For example, under the ULA model, we have
\begin{align}\label{eq:a_m far field}
a_{\text{I},m_{\text{I}}}(d\rightarrow \infty,\eta) \!= \! e^{-j\frac{2\pi}{\lambda} m_{\text{I}} d^{\text{I}}\cos\eta}, \forall m_I,
\end{align}which is the response in the classic far-field steering vector. Define 
\begin{align}
& d_k=\sqrt{(x_{\text{I}}-x_k)^2+(y_{\text{I}}-y_k)^2}, \forall k, \label{eqn:distance}\\
& \theta_k=\arctan \frac{y_{\text{I}}-y_k}{x_{\text{I}}-x_k}, ~ \forall k, \label{eqn:AOA}
\end{align}
as the distance and AOA from target $k$ to the IRS, respectively. Then, the channel of the user-target $k$-IRS path at coherence block $t$ is given by \cite{near_field_survey}
\begin{align}\label{eq:nlos channel k}
\boldsymbol{r}_{k,t} &= \beta_k \gamma_{k,t}\boldsymbol{a}_\text{I}(d_{k},\theta_k),\quad \forall k,t, 
\end{align}where $\beta_k$, $\forall k$, denotes the path loss factor of this path, and $\gamma_{k,t}$ is the radar cross section (RCS) of target $k$ in coherence block $t$. For example, under the Swerling target model \cite{swerling_model1}, $\gamma_{k,t}$'s are Gaussian distributed (their norms are Rayleigh distributed) independent over $k$ and $t$.
 
Next, the LOS MIMO channel between the IRS and the BS is given by
\vspace{-5pt}
\begin{align}\label{eq:los irs bs channel}
    \boldsymbol{\bar{G}} = \delta \boldsymbol{G},
\vspace{-5pt}
\end{align}where $\delta$ is the path-loss factor, and the element on the $m_\text{B}$-th row and $m_\text{I}$-th column of $\boldsymbol{G} \in \mathbb{C}^{M_{\text{B}} \times M_{\text{I}}}$ is denoted by \cite{near_field_survey}
\vspace{-5pt}
\begin{align}\label{eq:near irs bs channel}
    G_{m_{\text{B}},m_\text{I}} = e^{-j\frac{2\pi}{\lambda}d_{m_{\text{B}},m_\text{I}}^{\text{IB}}},\quad \forall m_{\text{B}},m_\text{I},
\end{align}with $d_{m_{\text{B}},m_{\text{I}}}^{\text{IB}}$ in meter being the distance between the $m_{\text{B}}$-th BS antenna and the $m_\text{I}$-th IRS element, $\forall m_{\text{B}},m_{\text{I}}$. In particular, when the BS is in the far-field region of the IRS, the LOS channel between the IRS and the BS can be simplified as
\begin{align}\label{equ:far_irs_bs_channel}
  \boldsymbol{\bar{G}} = \delta \boldsymbol{a}_{\rm B}(d \to \infty,\kappa) \boldsymbol{a}_{\rm I}^{\rm T}(d \to \infty,\xi),
\end{align}
where $\kappa$ and $\xi$ denote the AOA and angle of departure (AOD) of the LOS channel from the IRS to the BS, respectively. Since the locations of all BS antennas and IRS elements are known, we assume that $\boldsymbol{\bar{G}}$ is known in this paper.

\subsection{Problem Statement}

In this paper, our goal is to estimate the location $(x_k,y_k)$ of each target $k$ based on the BS received signals $\boldsymbol{\bar{\Upsilon}}_t^{(q)}$, $\forall q,t$, given in (\ref{eq:received signal}). Because there are no LOS links between targets and the BS, it is impossible to estimate the propagation delay and the AOA from each target to the BS based on $\boldsymbol{\bar{\Upsilon}}_t^{(q)}$, $\forall q,t$. Instead, we treat the IRS as a passive anchor and localize the targets by estimating their range and AOA information with regarding to the IRS, since LOS paths exist between targets and the IRS.

\section{Three-Phase Localization Protocol}\label{sec:protocol}

This paper adopts a three-phase protocol to exploit the IRS as a passive anchor for localizing the targets. In the first phase, we aim to estimate the CIRs $\boldsymbol{\bar{H}}_t^{(q)}$'s as given in (\ref{eq:multipath channel}) based on the signals received by the BS $\boldsymbol{\bar{\Upsilon}}_t^{(q)}$'s as given in (\ref{eq:received signal}). In the second phase, we aim to extract the range and AOA information of the targets with regarding to the IRS and/or the user from the estimated channels, since $\boldsymbol{\bar{H}}_t^{(q)}$'s are functions of delays and AOAs of the targets as shown in \eqref{eq:multipath channel l tap} and \eqref{eq:nlos channel k}. Specifically, we first detect whether there exist targets in each range cluster $l$. If a target is detected to be in range cluster $l$, then the propagation delay over the user-target-IRS-BS path is of $l$ OFDM samples. Next, given each range cluster $l$ with $\Omega_l \neq \emptyset$, we will design a MUSIC algorithm based approach on $\boldsymbol{\bar{h}}_{l,t}^{(q)}$, $\forall q, t$, to 1. detect the near-field and far-field targets; 2. estimate the AOA and the range from each near-field target to the IRS; and 3. estimate the AOA from each far-field target to the IRS. This is because $\boldsymbol{\bar{h}}_{l,t}^{(q)}$ can be viewed as a narrowband channel contributed by all the targets in range cluster $l$. In the third phase, by treating the IRS as a passive anchor, we aim to localize the targets based on the range and AOA information estimated in the second phase. In the following, we introduce each phase in details.

\subsection{Phase I: CIR Estimation}

Since this paper focuses on localizing static targets, the range cluster sets $\Omega_l$'s given in \eqref{eq:range cluster l} do not change with time. Based on \eqref{eq:multipath channel l tap}, given any $l$ with $\Omega_l = \emptyset$, it follows that the $l$-th row of $\boldsymbol{\bar{H}}_t^{(q)}$ given in \eqref{eq:multipath channel} is a zero vector, i.e., $\boldsymbol{\hat{h}}_{l,t}^{(q)} = \boldsymbol{0}, \forall q, t$. 
Therefore, we can exploit the above sparsity in all $\boldsymbol{\bar{H}}_t^{(q)}$'s to perform channel estimation over the $V$ coherence blocks jointly by solving the following group least absolute shrinkage and selection operator (LASSO) problem \cite{Yuan_2005_JRSSS}
\begin{align}\label{eq:group_lasso}
    \min_{\{\boldsymbol{\bar{H}}_t^{(q)}\}_{q=1,t=1}^{Q,V}}~ & \sum_{t=1}^{V}\sum_{q=1}^{Q} \left\| \boldsymbol{\bar{\Upsilon}}_t^{(q)} - \text{diag}(\boldsymbol{s}_t^{(q)})\boldsymbol{E}\boldsymbol{\bar{H}}_t^{(q)} \right\|_F^2 \notag \\
    & + \omega  \sum_{l=1}^{L}  \left(\sum_{q=1}^{Q}\sum_{t=1}^{V} \left\|\boldsymbol{\bar{h}}_{l,t}^{(q)}\right\|_2^2 \right)^{\frac{1}{2}}, 
\end{align}
where $\omega \geq 0$ is some given parameter to control the sparsity of the solution. The above problem (\ref{eq:group_lasso}) is a convex problem, which can be globally solved. Let $\boldsymbol{\tilde{H}}_t^{(q)}=[\boldsymbol{\tilde{h}}_{1,t}^{(q)},\cdots,\boldsymbol{\tilde{h}}_{L,t}^{(q)}]^T$ denote the optimal solution, $\forall q,t$.

\subsection{Phase II: Range and AOA Estimation for Near-Field and Far-Field Targets}\label{sec:phase_II}

In the second phase, we aim to estimate the target range and AOA information that is embedded in the estimated channels $\boldsymbol{\tilde{H}}_t^{(q)}$, $\forall q,t$. First of all, according to (\ref{eq:multipath channel l tap}), we declare that there are targets in range cluster $l$ if the estimated channels $\boldsymbol{\tilde{h}}_{l,t}^{(q)}$'s are strong, $\forall q,t$, and there are no targets in range cluster $l$ otherwise. Mathematically, the above detection is modeled as
\begin{align}
\hat{\Omega}_l\left\{\begin{array}{ll} \neq \emptyset, & {\rm if} ~ \sum\limits_{t=1}^V\sum\limits_{q=1}^Q \left\|\boldsymbol{\tilde{h}}_{l,t}^{(q)}\right\|^2\geq \rho_l, \\ =\emptyset, & \text{otherwsie}, \end{array}\right. ~ l=1,\dots,L, 
\end{align}where $\hat{\Omega}_l$ is an estimation of the set consisting of all the targets in range cluster $l$, i.e., $\Omega_l$ defined in (\ref{eq:range cluster l}), and $\rho_l>0$ is some pre-designed threshold. 
For convenience, define 
\begin{align}
    \Phi=\{l:\hat{\Omega}_l \neq \emptyset, \forall l\}
\end{align}
as the set consisting of the indices of all the range clusters with targets detected. For each $l \in \Phi$, the propagation delay from the user to all the targets in $\hat{\Omega}_l$ to the IRS to the BS is estimated to be of $l$ OFDM sample durations. Therefore, for each target in range cluster $l \in \Phi$, the range of the corresponding user-target-IRS-BS path is estimated as \cite{dvc}
\begin{align}\label{equ:d_UTIB}
    \hat{d}^{\rm UTIB}_l = \frac{(l-1)c_0}{N \Delta f} + \frac{1}{2N \Delta f} = \frac{(2l-1)c_0}{2N \Delta f}, \forall l,
\end{align}
where $c_0$ denotes the speed of the light. Moreover, based on the locations of the target $k$, the IRS, and the user, the range of the path from the user to target $k$ to the IRS and that from the IRS to the BS are
\begin{align}\label{eq:range user t irs}
    d^{\text{UTI}}(x_k,y_k) =&~ d^{\text{UT}}(x_k,y_k) + d^{\text{TI}}(x_k,y_k) \notag \\
    =&~ \sqrt{(x_{\text{I}}-x_k)^2 + (y_{\text{I}}-y_k)^2} \notag \\
    &~+\sqrt{(x_{\text{U}}-x_k)^2+(y_{\text{U}}-y_k)^2}, \forall k, \\
    d^{\rm IB} =&~ \!\sqrt{(x_{\text{I}}-x_{\text{B}})^2\!+\!(y_{\text{I}}-y_{\text{B}})^2}\!.
\end{align}
Therefore, for a target $k$ that is estimated in range cluster $l \in \Phi$, we have
\begin{align}\label{eqn:sum distance}
 d^{\text{UTI}}(x_k,y_k) &= \hat{d}^{\rm UTIB}_l - d^{\rm IB}+\!\mu_k, ~\forall k\in \hat{\Omega}_l,
\end{align}
where $\mu_k$ denotes the range estimation error for target $k$. In the case of perfect range estimation with $\mu_k=0$, (\ref{eqn:sum distance}) indicates that each target $k\in \hat{\Omega}_l$ should be on an ellipse because the sum of its distance to the IRS and that to the user is fixed. 

\begin{figure}[t]
   \centering
    \includegraphics[width=.4\textwidth]{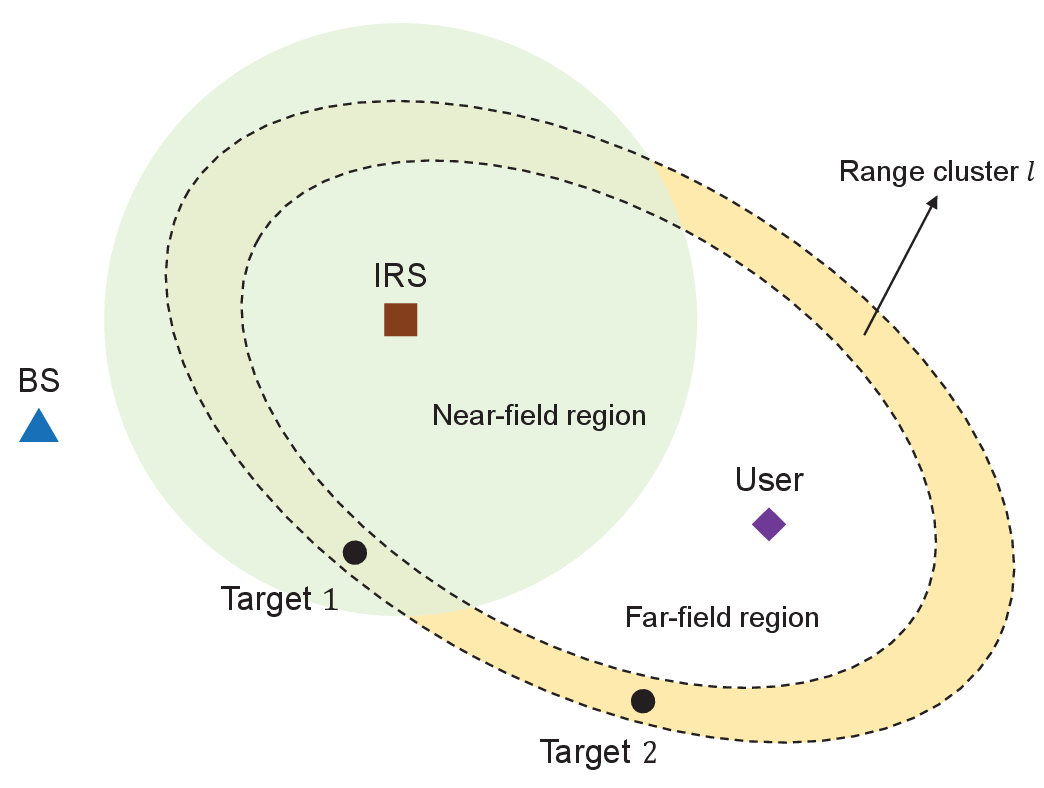}
    \vspace{-0.3cm}
    \caption{A toy example for coexistence of near-field and far-field targets, where target $1$ and target $2$ coexist within the range cluster $l$. They are in the different regions for the IRS. Specifically, target $1$ is in the near-field region of the IRS, while target $2$ is in the far-field region of the IRS.}\label{fig:demo_coexistence}
    \vspace{-0.5cm}
\end{figure}

After detecting the range clusters with targets in them, we have three goals listed as below:
\begin{itemize}
    \item {\bf{Goal $1$}}: Estimate the number of targets in each range cluster $l \in \Phi$, which is denoted by $\hat{K}_l = \left|\hat{\Omega}_l\right|$;
    \item {\bf{Goal $2$}}: Given each range cluster $l \in \Phi$, detect the set of targets that are in the far-field region of the IRS, denoted by the set $\hat{\Omega}^{\rm F}_{l}$, and the set of targets that are in the near-field region of the IRS, denoted by the set of $\hat{\Omega}^{\rm N}_{l}$;\footnote{One example is given in Fig. \ref{fig:demo_coexistence} to show that even in the same range cluster, it is likely that some users are in the near-field region of the IRS, while others are in the far-field region.}
    \item {\bf{Goal $3$}}: If a target $k$ is detected to be a far-field target, then we estimate its AOA to the IRS, denoted by $\hat{\theta}_{k}$, based on the relation between the AOA and the far-field steering vector $\boldsymbol{a}_\text{I}(d\to\infty,\theta)$; otherwise, we estimate both its AOA and distance to the IRS, denoted by $\hat{d}_k$ and $\hat{\theta}_{k}$, respectively, based on the relation among the AOA, the distance, and the near-field steering vector $\boldsymbol{a}_\text{I}(d,\theta)$.
\end{itemize}

Because it takes quite large space to show how to achieve \textbf{Goals} 1-3, in the rest of this section, we will assume that the target number $\hat{K}_l$, far-field target set $\hat{\Omega}^{\rm F}_{l}$, near-field target set $\hat{\Omega}^{\rm N}_{l}$, AOAs $\hat{\theta}_{k}$'s for far-field targets, AOA and range pairs $(\hat{d}_k,\hat{\theta}_k)$ for near-field targets have all been successfully estimated and merely introduce how to localize the targets based on these estimations in Phase III. This may help readers quickly understand the overall process of our proposed three-phase protocol. We will show how to achieve the above three goals of Phase II in Section \ref{sec:rank_two_case}.

\subsection{Phase III: Localization}

In the third phase, we aim to estimate the locations of the targets, i.e., $(x_k,y_k)$'s, $\forall k$, assuming that \textbf{Goals} 1-3 in Phase II listed in the above have been achieved. 

First, we show how to localize the users that are in the far-field region of the IRS. Given each far-field target $k \in \hat{\Omega}_l^{\rm F}$ with some $l \in \Phi$, the range of the path from the user to it to the IRS to the BS is estimated as $\hat{d}^{\rm UTI}$ according to \eqref{equ:d_UTIB}, and its AOA to the IRS is estimated as $\hat{\theta}_k$ in Phase II. Then, the weighted residual minimization problem to localize the far-field target $k \in \hat{\Omega}_{l}^{\rm F}$ can be formulated as
\begin{small}
\begin{align}\label{equ:loc_far_target_k}
    \min_{(x_k,y_k)} &~ (1-\varpi_{\rm F}) \left( d^{\rm UTI}(x_k,y_k) + d^{\rm IB} - \hat{d}^{\rm UTIB}_l \right)^2 \notag  \\
    &~+ \varpi_{\rm F} \left( \arctan\frac{y_{\rm I} - y_k}{x_{\rm I} - x_k} - \hat{\theta}_{k} \right)^2, 
\end{align}
\end{small}where $\varpi_{\rm F} \in [0,1]$ is the tuning parameter to balance the angle estimation residual and the range estimation residual of a far-field target. Then, the unique solution to problem \eqref{equ:loc_far_target_k} can be considered as the estimation of the location of this far-field target $k$. 
It can be show that the following solution (consider targets are only located in one side of the IRS) leads to zero-residual to the above problem
\begin{small}
\begin{subequations}
\begin{align}
    \hat{x}_k =& \frac{ \left(\frac{(2l-1) c_0}{2N\Delta f} - d^{\rm IB} \right)^2 - (x_{\rm I} - x_{\rm U})^2 - (y_{\rm I} - y_{\rm U})^2}{2\left(x_{\rm I} - x_{\rm U} + (y_{\rm I} - y_{\rm U})\tan\hat{\theta}_k - \frac{\frac{(2l-1) c_0}{2N\Delta f} - d^{\rm IB}}{\cos\hat{\theta}_k}\right)} + a_{\rm I},  \\
    \hat{y}_k =&  \frac{ \left(\frac{(2l-1) c_0}{2N\Delta f} - d^{\rm IB} \right)^2 - (x_{\rm I} - x_{\rm U})^2 - (y_{\rm I} - y_{\rm U})^2}{2\left(x_{\rm I} - x_{\rm U} + (y_{\rm I} - y_{\rm U})\tan\hat{\theta}_k - \frac{\frac{(2l-1) c_0}{2N\Delta f} - d^{\rm IB}}{\cos\hat{\theta}_k}\right)} \tan\hat{\theta}_k + b_{\rm I}.
\end{align}
\end{subequations}
\end{small}Therefore, the above solution is optimal.

Second, we show how to localize the targets that are in the near-field region of the IRS. Given each near-field target $k \in \hat{\Omega}_{l}^{\rm N}$ with some $l \in \Phi$, the range of the path from the target to the IRS is estimated as $\hat{d}_{k}$, and its AOA to the IRS is estimated as $\hat{\theta}_k$ in Phase II. We can then formulate the weighted residue minimization problem to localize the near-field target $k \in \hat{\Omega}_{l}^{\rm N}$ as 
\begin{small}
\begin{align}\label{equ:loc_near_target_k}
    \min_{(x_k,y_k)} &~ \varpi_{{\rm N},1}  \left( \arctan\frac{y_{\rm I} - y_k}{x_{\rm I} -  x_k} - \hat{\theta}_{k} \right)^2  \notag \\
    &~+ \varpi_{{\rm N},2} \left( d^{\rm UTI}(x_k,y_k) + d^{\rm IB} - \frac{(2l-1) c_0}{2N\Delta f} \right)^2  \notag \\
    &~+ (1 - \varpi_{{\rm N},1} - \varpi_{{\rm N},2})\left( d^{\rm TI}(x_k,y_k) - \hat{d}_k \right)^2, 
\end{align}
\end{small}where $\varpi_{{\rm N},1} \in [0,1]$ and $\varpi_{{\rm N},2} \in [0,1]$ with $\varpi_{{\rm N},1} + \varpi_{{\rm N},2} \in [0,1]$ are the tuning parameters to balance the estimation error by Algorithm \ref{alg:pri_music} and the range estimation error introduced by the sum distance estimation in Phase I. 
In contrast to the far-field target localization, it is hard to get the closed-form solution for the problem \eqref{equ:loc_near_target_k}, and we can apply Gauss-Newton algorithm \cite{Torrieri_1984_TAES} to solve it.
\section{Achieving Goals 1-3 in Phase II}\label{sec:rank_two_case}

In this section, we give the specific schemes to realize \textbf{Goals} 1-3 in Phase II listed in Section \ref{sec:phase_II} to complete our three-phase protocol. 


For illustration, we define $\Omega_l^\text{F}$ and $\Omega_l^\text{N}$ as the ture sets of far-field targets and near-field targets in range cluster $l \in \Phi$, respectively, such that $\hat{\Omega}_{l}^{\rm F}$ and $\hat{\Omega}_{l}^{\rm N}$ in \textbf{Goal} 2 are their estimations.
According to \eqref{eq:cascaded channel}, \eqref{eq:nlos channel k}, and \eqref{eq:los irs bs channel}, the effective channels of all targets in range cluster $l \in \Phi$ that are estimated via problem \eqref{eq:group_lasso} satisfy
\begin{subequations}\label{eq:estimated_channel cluster l}
\begin{align}%
    \boldsymbol{\tilde{h}}_{l,t}^{(q)} &= \boldsymbol{\bar{h}}_{l,t}^{(q)}+\boldsymbol{\tilde{z}}_{l,t}^{(q)} \label{equ:noiseless+noise} \\
    &=\sum_{k\in\Omega_l}\delta\boldsymbol{G}\text{diag}(\boldsymbol{\phi}_t^{(q)}) \boldsymbol{r}_{k,t} + \boldsymbol{\tilde{z}}_{l,t}^{(q)} \label{equ:expand_estimated_channel} \\
    &= \delta\boldsymbol{G}\text{diag}(\boldsymbol{\phi}_t^{(q)}) \Big(\sum_{k\in\Omega_l^\text{F}} \beta_{k} \gamma_{k,t} \boldsymbol{a}_\text{I}(d_{k} \to \infty,\theta_{k}) \notag  \\
    &~\quad + \sum_{k\in\Omega_l^\text{N}} \beta_{k} \gamma_{k,t} \boldsymbol{a}_\text{I}(d_{k},\theta_{k})\Big) + \boldsymbol{\tilde{z}}_{l,t}^{(q)} \label{equ:expand_far_near}  \\
    &= \sum_{k\in\Omega_l} v_{k,t} \boldsymbol{\psi}^{(q)}_t(\bar{d}_k,\theta_k) + \boldsymbol{\tilde{z}}_{l,t}^{(q)} \label{equ:weighted_steering} \\
    &= \boldsymbol{\Psi}_{t}^{(q)}(\Theta_{l}) \boldsymbol{\upsilon}_{l,t} + \boldsymbol{\tilde{z}}_{l,t}^{(q)},  \quad\forall t, l \in \Phi. \label{equ:estimate_h_matrix_form} 
\end{align}
\end{subequations}
In the above, $\boldsymbol{\tilde{z}}_{l,t}^{(q)}$ in \eqref{equ:noiseless+noise} is the error of estimating $\bar{\boldsymbol{h}}_{l,t}^{(q)}$; 
$\boldsymbol{\psi}_t^{(q)}(\bar{d}_{k},\theta_{k})$ in \eqref{equ:weighted_steering} is defined as
\begin{align}\label{equ:effective_steering}
    \boldsymbol{\psi}_t^{(q)}(\bar{d}_{k},\theta_{k}) = \boldsymbol{G}\text{diag} (\boldsymbol{\phi}_t^{(q)}) \boldsymbol{a}_\text{I}(\bar{d}_{k},\theta_k), \forall k \in \Omega_l,
\end{align}
where $\bar{d}_k$ is the effective distance of target $k$ to the IRS, which is defined as
\begin{align}
\bar{d}_{k}=\left\{\begin{matrix}
 d_{k}, & \text{if}~k\in\Omega_l^{\rm N},  \\ 
 \infty, & \text{if}~k\in\Omega_l^{\rm F};
\end{matrix}\right.\label{eq:model range}
\end{align}
$\boldsymbol{\Psi}_{t}^{(q)}(\Theta_{l}) = \boldsymbol{G}\text{diag} (\boldsymbol{\phi}_t^{(q)})\boldsymbol{A}_{\rm I}(\Theta_l) \in \mathbb{C}^{M_{\rm B} \times K_l}$ in \eqref{equ:estimate_h_matrix_form} is defined as the effective steering matrix and $\boldsymbol{A}_{\rm I}(\Theta_l) \in \mathbb{C}^{M_{\rm I} \times K_l}$ is the steering matrix for the target-IRS channels with each column given by $\boldsymbol{a}_{\rm I}(\bar{d}_k,\theta_k), ~\forall k \in \Omega_l$, where 
$\Theta_l = \{ (\bar{d}_k,\theta_{k}) | k \in \Omega_l \}$ is the set consisting of the pairs of the effective distances and AOAs of all targets located in the range cluster $l \in \Phi$;
and $\boldsymbol{v}_{l,t} \in \mathbb{C}^{K_l \times 1}$ is the vector consisting of $v_{k,t}$'s, $\forall k \in \Omega_l$, in range cluster $l$. Note that \eqref{equ:expand_far_near} holds because the steering vector of a far-field target can be expressed as $\boldsymbol{a}_\text{I}(d\to\infty,\theta)$ as shown in Section \ref{subsec:channel model}. 




As mentioned before, one key challenge to leverage the IRS as a passive anchor is the passive nature of the IRS - we need to estimate the AOAs and/or ranges from the targets to the IRS based on the signals received by the BS. Amazingly, by treating $\boldsymbol{\psi}_{t}^{(q)}(\bar{d}_k,\theta_k)$ in \eqref{equ:effective_steering} as the steering vector of the IRS towards the target $k$, whose AOA and effective distance to the IRS are denoted by $\theta_k$ in \eqref{eqn:AOA} and $\bar{d}_{k}$ in \eqref{eq:model range}, equation \eqref{eq:estimated_channel cluster l} mathematically describes a virtual system where the effective observations, $\boldsymbol{\tilde{h}}_{l,t}^{(q)}$'s, are the weighted sum of the IRS's effective steering vectors towards the targets in the range cluster $l$. 
Such a virtual system is similar to the conventional multi-antenna system for estimating AOA and/or range information of the near-field and far-field targets, where there are LOS paths between targets and the multi-antenna receive anchor, and the signals received by the anchor is the weighted sum of steering vectors towards the targets \cite{Zheng_2019_TAP}. Therefore, our localization task is to extract the ranges and AOAs from the estimated channels $\{\boldsymbol{\tilde{h}}_{l,t}^{(q)}\}_{q=1,t=1}^{Q,V}$. In the literature, there exist plenty of classic signal processing methods to deal with the system described in \eqref{eq:estimated_channel cluster l}, such as the MUSIC algorithm and the ESPRIT algorithm \cite{aoa_survey}. In this paper, we will apply the MUSIC algorithm to extract the AOA information of the far-field targets and the AOA together with the range information of the near-field targets from $\{\boldsymbol{\tilde{h}}_{l,t}^{(q)}\}_{q=1,t=1}^{Q,V}$. 

However, there are two differences between our AOA and/or range estimation system \eqref{eq:estimated_channel cluster l} and the conventional multi-antenna localization system \cite{Zheng_2019_TAP} that make it impossible to directly apply the MUSIC algorithm on \eqref{eq:estimated_channel cluster l}.

\begin{itemize}
\item \textbf{Difference 1}: In the conventional localization system, the steering vectors are static over time. However, in our system \eqref{eq:estimated_channel cluster l}, because the IRS reflecting coefficients $\boldsymbol{\phi}_t^{(q)}$'s can change over different OFDM symbol durations and different coherence blocks, the steering vectors can be dynamic. 

\item \textbf{Difference 2}: In the conventional localization system, it is required that the number of receive antennas is larger than the number of targets. Mathematically, this implies that the rank of the matrix that contains the steering vectors towards all the targets is equal to the number of targets such that the null space of this matrix exists to implement the MUSIC algorithm. However, the rank of $\boldsymbol{\Psi}_{t}^{(q)}(\Theta_{l})$ in \eqref{equ:estimate_h_matrix_form} can be smaller than the number of targets, even when the number of BS antennas and the number of IRS reflecting elements are both larger than the number of targets in each range cluster. This is because the rank of $\boldsymbol{\Psi}_{t}^{(q)}(\Theta_{l}) = \boldsymbol{G} \text{diag} (\boldsymbol{\phi}^{(q)}) \boldsymbol{A}_{\rm I}(\Theta_l)$ is limited by the rank of $\boldsymbol{G}$. For example, when the BS and the IRS are in the far-field region of each, it can be shown that the rank of the LOS channel between the BS and the IRS, which is denoted by $r_{\boldsymbol{G}}$, is one. In this case, the rank of $\boldsymbol{\Psi}_{t}^{(q)}(\Theta_{l})$ is also one, and it is impossible to estimate the AOA and/or range information of the targets in each range cluster $l \in \Phi$ based on the MUSIC algorithm.
\end{itemize}

In the following, we propose a novel method to tackle the above two challenges for enabling the MUSIC algorithm on \eqref{eq:estimated_channel cluster l}. There are two key ideas under our proposed method. First, given any $q$th OFDM symbol duration, we set a common IRS reflecting pattern, denoted by $\boldsymbol{\bar{\phi}}^{(q)}$, over all the coherence blocks, i.e., 
\begin{align}\label{equ:phi_equ_t} 
    \boldsymbol{\phi}_{t}^{(q)}=\boldsymbol{\bar{\phi}}^{(q)},\quad \forall t, q.
\end{align}
In this case, the effective steering vectors in (\ref{equ:effective_steering}) reduce to
\begin{align}\label{eq:new array}
    \boldsymbol{\bar{\psi}}^{(q)}(\bar{d}_{k},\theta_{k}) = \boldsymbol{G}\text{diag}(\boldsymbol{\bar{\phi}}^{(q)}) \boldsymbol{a}_\text{I}(\bar{d}_{k},\theta_{k}), \forall q,t, k \in \Omega_l,
\end{align}
and the estimated channels given in (\ref{eq:estimated_channel cluster l}) reduce to
\begin{align}\label{equ:est_channel_reduce}
  \boldsymbol{\tilde{h}}_{l,t}^{(q)} &= \sum_{k\in\Omega_l} v_{k,t} \boldsymbol{\bar{\psi}}^{(q)}(\bar{d}_k,\theta_k) + \boldsymbol{\tilde{z}}_{l,t}^{(q)} \notag \\
  &= \bar{\boldsymbol{\Psi}}^{(q)}(\Theta_{l}) \boldsymbol{\upsilon}_{l,t} + \boldsymbol{\tilde{z}}_{l,t}^{(q)}, ~\forall t, l\in \Phi,
\end{align}
where $\bar{\boldsymbol{\Psi}}^{(q)}(\Theta_{l})  \in \mathbb{C}^{M_{\rm B} \times K_l}$ has similar definition to $\boldsymbol{\Psi}_{t}^{(q)}(\Theta_{l})$.

\begin{figure}[t]
	\centering
    \includegraphics[width=.47\textwidth]{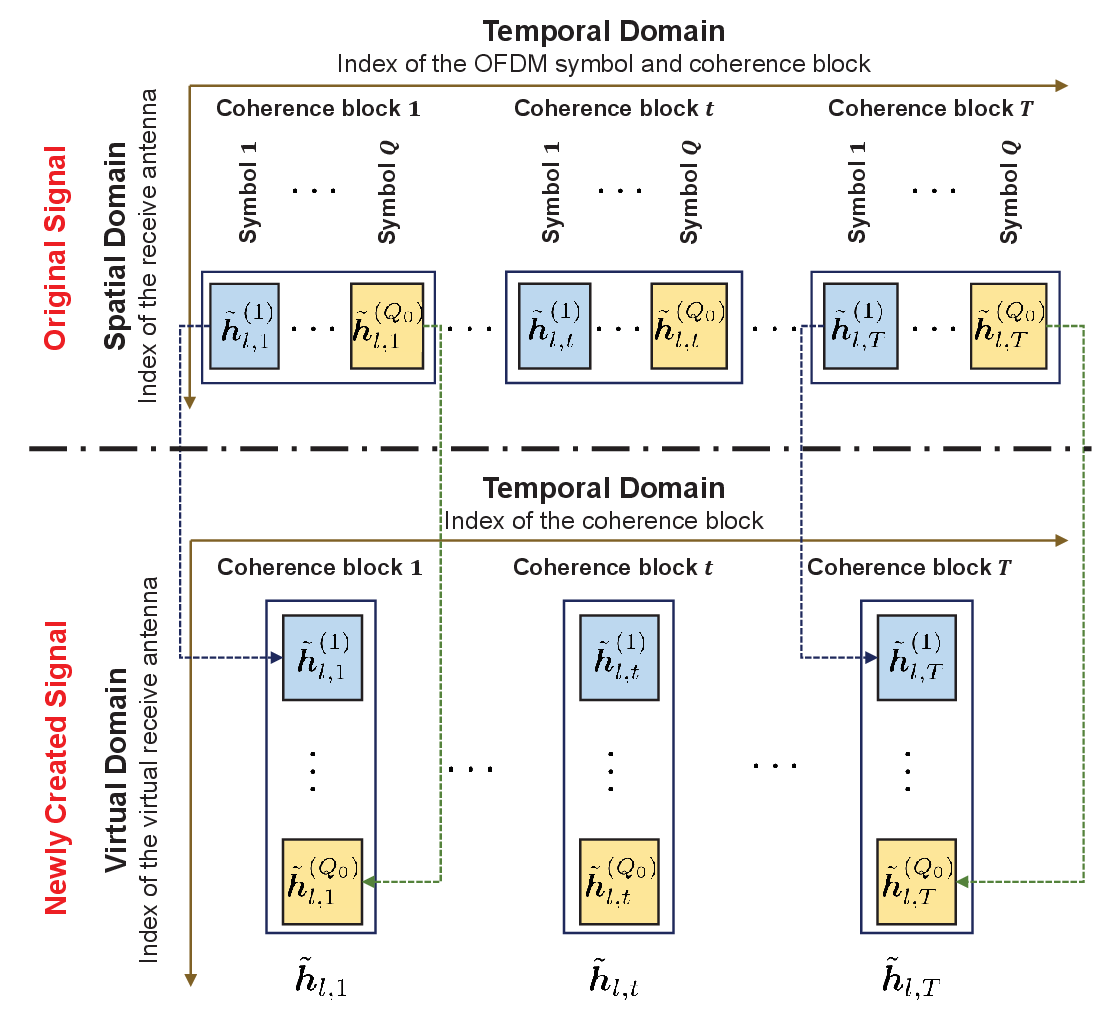}
    \vspace{-0.3cm}
    \caption{Illustration of original signal model and newly created signal model. Specifically, we combine all the estimated channels of the first $Q_0$ OFDM symbols within one coherence block together to create a new virtual high-dimension signal, which contains information in both spatial (BS antennas) and temporal (OFDM symbols) domain.}\label{fig:multisignal}
    \vspace{-0.6cm}
\end{figure}

After adopting the first idea to employ a common IRS reflecting pattern at the same OFDM symbol durations of different coherence blocks in \eqref{equ:phi_equ_t}, the two challenges listed in the above still exist in the new system described in \eqref{eq:estimated_channel cluster l}. In the following, we introduce the second idea that can tackle these two challenges together based on \eqref{equ:est_channel_reduce} - leveraging temporal domain signals. Specifically, at each coherence block $t$, given any range cluster $l \in \Phi$, we combine the temporal domain channels over the first $Q_0 \le Q$ OFDM symbol durations together to create a new virtual channel vector as follows
\begin{align}\label{equ:virtual_channel_l}
    \boldsymbol{\tilde{h}}_{l,t} = [(\boldsymbol{\tilde{h}}_{l,t}^{(1)})^T,\dots,(\boldsymbol{\tilde{h}}_{l,t}^{(Q_0)})^T]^T, ~\forall t, l \in \Phi.
\end{align}
The selection of $Q_0$ will be discussed later.
According to \eqref{equ:est_channel_reduce}, it can be shown that
\begin{align}
    \boldsymbol{\tilde{h}}_{l,t} &= \sum_{k\in\Omega_l} v_{k,t} \boldsymbol{\breve{\psi}}(\bar{d}_k,\theta_k) + \boldsymbol{\tilde{z}}_{l,t} \notag \\ 
    &= \boldsymbol{\breve{\Psi}}(\Theta_{l}) \boldsymbol{\upsilon}_{l,t}  + \boldsymbol{\tilde{z}}_{l,t}, ~\forall t, l \in \Phi,
    \label{equ:estimated_channel_deficient_noisy}
\end{align}
where
\begin{align}
    \boldsymbol{\breve{\Psi}}(\Theta_{l}) &= \boldsymbol{P}\boldsymbol{A}_{\rm I}(\Theta_{l}), ~\forall l \in \Phi, \label{equ:vir_steering} \\ 
    \boldsymbol{P} &= [(\boldsymbol{P}^{(1)})^T,\dots,(\boldsymbol{P}^{(Q_0)})^T]^T, \\
    \boldsymbol{P}^{(q)} &= \boldsymbol{G}\text{diag}(\boldsymbol{\bar{\phi}}^{(q)}), ~\forall q \in \{1,\dots,Q_0\},\\
    \boldsymbol{\tilde{z}}_{l,t} &= [(\boldsymbol{\tilde{z}}_{l,t}^{(1)})^T,\dots,(\boldsymbol{\tilde{z}}_{l,t}^{(Q_0)})^T]^T, ~\forall t, l \in \Phi. \label{equ:vir_noise}
\end{align}

The above creation of the new virtual signals by combining the signals in both the spatial and temporal domains is illustrated in Fig. \ref{fig:multisignal}.
Next, we show that the two challenges listed in the above disappear in the new system described in \eqref{equ:estimated_channel_deficient_noisy} such that we can apply the MUSIC algorithm to estimate the AOA and/or range information of all the targets in each range cluster $l \in \Phi$. First, it is observed that the virtual steering matrices $\boldsymbol{\breve{\Psi}}(\Theta_{l})$'s in \eqref{equ:vir_steering} are static over time. Second, we tackle the challenge about matrix rank. 
For convenience, we define $K^{\rm max} = \max\{K_l, \forall l \in \Phi\}$
\begin{theorem}\label{theorem1}
We have $\text{rank}(\breve{\boldsymbol{\Psi}}(\Theta_{l})) = K_l$, $\forall l \in \Phi$ almost surely when the following conditions are satisfied:
\begin{enumerate}
        \item The number of utilized OFDM symbols satisfies $Q_0 \ge K^{\rm max}$;
        \item The IRS reflecting elements follow ULA with element spacing $\frac{\lambda}{2}$ and the number of IRS elements satisfies $M_{\rm I} \ge K^{\rm max}$;
        \item The coefficient of each reflecting element $m_{\rm I}$ during OFDM symbol $q$ is set as
        \vspace{-6pt}
        \begin{align}\label{equ:IRS_phi_design}
            &\bar{\phi}^{(q)}_{m_{\rm I}} = \frac{w_{m_{\rm I},q}^{\rm I}}{G_{1,m_{\rm I}}}e^{j (m_{\rm I}-1) \vartheta}, \notag \\
            & \forall m_{\rm I} \in \{1,\dots,M_{\rm I}\}, q \in \{1,\dots,Q_0\},
        \end{align}
        where $w_{m_{\rm I},q}^{\rm I}$ is the element on the $m_{\rm I}$-th row and the $q$-th column element of the DFT matrix $\boldsymbol{W}^{\rm I} \in \mathbb{C}^{M_{\rm I} \times M_{\rm I}}$ and $\vartheta$ is an arbitrary angle in the region of $[0, 2\pi)$.
\end{enumerate}
\end{theorem}

\begin{IEEEproof}
Please refer to Appendix \ref{appendix1}.
\end{IEEEproof}

The first condition can be easily satisfied since the number of OFDM symbols in each coherence block is much larger than $K^{\rm max}$ in practice. Moreover, in practice, the IRS consists of a huge number of reflecting elements, and a common spacing of $\lambda/2$ between adjacent elements is a standard IRS pattern.\footnote{Although we cannot prove this rigorously, we find in all the numerical examples that Theorem 1 is still true if the IRS element spacing is less than $\lambda/2$.} Thus, the second condition also holds. 
Theorem 1 thus implies that as long as the IRS reflecting coefficients are set as \eqref{equ:IRS_phi_design}, then the MUSIC algorithm can work on the signals \eqref{equ:virtual_channel_l} almost surely. Note that if $Q_0=1$, i.e., $\boldsymbol{\tilde{h}}_{l,t} = \boldsymbol{\tilde{h}}^{(1)}_{l,t}, ~\forall t, l \in \Phi$, we do not combine channels over different symbols as shown in \eqref{equ:virtual_channel_l}, then the rank of $\breve{\boldsymbol{\Psi}}(\Theta_{l})$ is limited by the rank of the BS-IRS channel, i.e., $r_{\boldsymbol{G}}$. In this case, the MUSIC algorithm cannot work on the signals given in \eqref{equ:estimated_channel_deficient_noisy} if $r_{\boldsymbol{G}}< K^{\rm max}$. Theorem \ref{theorem1} shows that by creating a virtual signal as in \eqref{equ:virtual_channel_l} that consists of channel vectors over a sufficient number of OFDM symbols, the MUSIC algorithm can work no matter what is the value of $r_{\boldsymbol{G}}$. 

One more comment is about the first condition in Theorem \ref{theorem1}. Although we cannot prove it rigorously, we find via a vast number of numerical examples that $\text{rank}(\breve{\boldsymbol{\Psi}}(\Theta_{l})) = K_l, \forall l \in \Phi$ even when the first condition in Theorem \ref{theorem1} is relaxed to $Q_0 r_{\boldsymbol{G}} \ge K^{\rm max}$. Intuitively, this is because $\text{rank}(\boldsymbol{P})=Q_0 r_{\boldsymbol{G}}$ and according to \eqref{equ:vir_steering}, the rank of $\breve{\boldsymbol{\Psi}}(\Theta_{l}) = \boldsymbol{P} \boldsymbol{A}(\mathcal{D}_l)$ should be upper bounded by $Q_0 r_{\boldsymbol{G}}$. Note that a small $Q_0$ is preferred in practice because the complexity to implement the MUSIC algorithm to \eqref{equ:estimated_channel_deficient_noisy} depends on the dimension of the virtual signals. Therefore, in practice, we can set $Q_0=\ceil*{\frac{K^{\rm max}}{r_{\boldsymbol{G}}}}$ to make the MUSIC algorithm applicable to \eqref{equ:estimated_channel_deficient_noisy} with the minimum complexity.  
Based on Theorem \ref{theorem1}, we can have the following corollary.
\begin{corollary}
In the asymptotic regime with $V \to \infty$, both of the far-field and near-field targets can be perfectly detected by the MUSIC algorithm when the three conditions in Theorem \ref{theorem1} are satisfied. In particular, the AOA information of the far-field targets as well as the AOA and range information of the near-field targets can be perfectly estimated. 
\end{corollary}
\begin{IEEEproof}
In the asymptotic regime with $V \to \infty$, the sample covariance matrix of the virtual channel $\tilde{\boldsymbol{h}}_{l,t}$ is equal to the true covariance matrix. Then, the virtual steering vectors $\boldsymbol{\breve{\psi}}(\bar{d}_k,\theta_k)$'s of all targets can be perfectly detected by the MUSIC algorithm \cite{Wax_1989_TASSP} when all conditions in Theorem \ref{theorem1} hold. Since there is a one-to-one mapping between the pair $(\bar{d}_k,\theta_k)$ and $\boldsymbol{\breve{\psi}}(\bar{d}_k,\theta_k)$, the effective distances and AOAs of all targets can also be perfectly derived. From \eqref{eq:model range}, we can perfectly detect which target is in the far-field region and which target is in the near-field region. Due the fact that $\bar{d}_k \to \infty$ for far-field targets by \eqref{eq:model range}, only the AOA information can be perfectly estimated for each far-field target, while both of the AOA and range information for each near-field target can be perfectly estimated.
\end{IEEEproof}

In the following, we show how to achieve \textbf{Goals} 1-3 listed in Section \ref{sec:phase_II} via applying the MUSIC algorithm on the system described in \eqref{equ:estimated_channel_deficient_noisy}.

First, we consider the target number estimation problem of \textbf{Goal} $1$, which is referred as the standard model selection problem \cite{Akaike_1974_TAC}.
Then, we propose to utilize the commonly adopted Akaike information criterion (AIC) criterion \cite{Akaike_1974_TAC} to solve this problem. In detail, we first calculate the sample covariance matrix of the virtual channel in \eqref{equ:estimated_channel_deficient_noisy} as
\begin{align}\label{equ:sampled_cov_vs_h}
    \boldsymbol{\tilde{R}}_{l} &= \frac{1}{V} \sum_{t=1}^{V} \boldsymbol{\tilde{h}}_{l,t} \boldsymbol{\tilde{h}}_{l,t}^H, ~\forall l \in \Phi.
\end{align}
By performing EVD, we can derive $\boldsymbol{\tilde{R}}_{l} = \boldsymbol{U}_l \boldsymbol{\it \Lambda}_{l} \left( \boldsymbol{U}_l \right)^{H}$, where $\boldsymbol{U}_l = [\boldsymbol{u}_{l,1},\dots,\boldsymbol{u}_{l,Q_0 M_{\rm B}}]$ is the eigenvector matrix and $\boldsymbol{\it \Lambda}_{l} = \text{diag}([\lambda_{l,1},\dots,\lambda_{l,Q_0 M_{\rm B}}] )$ is the eigenvalue matrix with $\lambda_{l,1} \ge \lambda_{l,2} \ge \dots \ge \lambda_{l,Q_0 M_{\rm B}}$. Then the estimated target number $\hat{K}_l$ based on the AIC criterion is given by
\begin{small}
\begin{align}\label{eq:est k}
    \hat{K}_l = &\arg\max_{K_l} ~\log\left(\frac{\prod_{i=K_l+1}^{Q_0 M_\text{B}}\lambda_{l,i}^{\frac{1}{M_\text{B}-K_l}}}{\frac{1}{M_\text{B}-K_l}\sum_{j=K_l+1}^{Q_0 M_\text{B}}\lambda_{l,j}}\right)^{Q_0 M_\text{B}-K_l} \notag \\
    & \qquad\qquad - 2K_l(Q_0 M_\text{B} - K_l),\quad \forall l \in \Phi.
\end{align}
\end{small}Note that $\hat{K}_l$ can be easily obtained via exhaustive search.

Then we consider to realize {\bf{Goal $2$}} and {\bf{Goal $3$}} jointly.
Define $\boldsymbol{\bar{U}}_l =[\boldsymbol{u}_{l,\hat{K}_l+1},\cdots,\boldsymbol{u}_{l,Q_0 M_\text{B}}]\in\mathbb{C}^{Q_0 M_\text{B} \times (Q_0 M_\text{B}-\hat{K}_l)}, \forall l \in \Phi$. Note that the key difference between a near-field target and a far-field target lies in the steering vector - the effective distance of target $k$, i.e., $\bar{d}_k$, goes to infinity if $k\in\Omega_l^\text{F}$, and is finite if $k\in\Omega_l^\text{N}$. As a result, we may use this difference to detect whether a target is a far-field target or a near-field target, and estimate its AOA and range information to the IRS. 
Specifically, for near-field targets in range cluster $l \in \Phi$, the MUSIC algorithm require us to calculate the following 2D spectrum
\begin{small}
\begin{align}\label{eq:spectrum 2D}
     P_l^{\text{N}}(d,\theta) =&~\frac{\boldsymbol{\breve{\psi}}(d,\theta)^H\boldsymbol{\breve{\psi}}(d,\theta)}{\boldsymbol{\breve{\psi}}(d,\theta)^H\boldsymbol{\bar{U}}_l\boldsymbol{\bar{U}}_l^H\boldsymbol{\breve{\psi}}(d,\theta)}, \theta \in [0,\pi),d \in (0,\infty), 
\end{align}
\end{small}where $\boldsymbol{\breve{\psi}}(d,\theta)$ is similarly defined as each column vector in \eqref{equ:vir_steering}.
Define $d_{\rm max}=L c_0/N\Delta f - d^{\rm IR}$ as the maximum distance that can be detected in the $L$-tap environment. Then, given the step size $\Delta d$ for searching $d$ and step size $\Delta {\theta}$ for searching $\theta$, define 
\begin{small}
\begin{align}\label{equ:R_dtheta}
  \mathcal{R}^{\rm N}=\bigg\{(d,\theta)| &d = \zeta\Delta{d}, \zeta = 1,\dots,\ceil*{\frac{d_{\rm max}}{\Delta d}}, \notag \\
                           &\theta = \mu\Delta{\theta}, \mu = 1,\dots,\ceil*{\frac{\pi}{\Delta {\theta}}} \bigg\} \cap \mathcal{L}^{\rm N},
\end{align}
\end{small}as the set consisting of all discrete grids in the search region of $d$ and $\theta$, where $\mathcal{L}^{\rm N}$ is the near-field region for the IRS and can be derived in advance. Over $\mathcal{R}^{\rm N}$, we can perform a 2D exhaustive search to find the peaks of $P_l^{\rm N}(d,\theta), \forall l\in \Phi$. Define $\Xi_l^{\rm N}$ as the set of all pairs of $(d,\theta)$ that lead to peaks of $P_l^{\rm N}(d,\theta)$. Then, the pairs of distance and AOA of all the near-field targets in range cluster $l \in \Phi$ are contained in $\Xi_l^{\rm N}$.

Next, we consider the detection of far-field targets in the range cluster $l \in \Phi$. Define the 1D spectrum as
\begin{small}
\begin{align}\label{eq:spectrum 1D}
     P_l^{\text{F}}(\theta) 
     =&~ \frac{\boldsymbol{\breve{\psi}}(d\to\infty,\theta)^H\boldsymbol{\breve{\psi}}(d\to\infty,\theta)}{\boldsymbol{\breve{\psi}}(d\to\infty,\theta)^H \bar{\boldsymbol{U}}_l\boldsymbol{\bar{U}}_l^H\boldsymbol{\breve{\psi}}(d\to\infty,\theta)}, \theta\in [0,\pi). 
\end{align}
\end{small}Since the effective distance of far-field targets are considered to be infinity, we only perform a 1D angle search over $P_l^{\text{F}}(\theta)$ to find its peaks. 
The set of search grids for far-field targets can be given as
\begin{align}\label{equ:R_theta}
    \mathcal{R}^{\rm F} = \left\{ \theta | \theta = \xi\Delta{\theta}, \xi = 1,\dots,\ceil*{\frac{\pi}{\Delta {\theta}}} \right\}.
\end{align}
Define $\Xi_l^\text{F}$ as the AOA set that leads to peaks of $P_l^{\text{F}}(\theta), \forall l \in \Phi$.


However, to realize precise estimation on the AOAs and/or ranges of targets, the selected step sizes $\Delta d$ and $\Delta {\theta}$ is usually much smaller than $d_{\rm max}$ and $\pi$, respectively. So that the number of seach grids in $\mathcal{R}^{\rm N}$ is quite large and it is of prohibitive complexity to calculate the 2D spectrum, as required by the MUSIC algorithm. In fact, we can leverage the prior information of the range cluster to help reduce the search region, leading to much lower searching complexity. Such method is named as the prior information-assisted MUSIC algorithm. Specifically, if targets considered to be in the range cluster $l$, their positions $(x_k, y_k)$'s must satisfy the following condition
\begin{align} \label{equ:pri_inf_cluster_l}
    \frac{(l-1)c_0}{N \Delta f} \le d^{\rm UTI}(x_k,y_k) + d^{\rm IB}  < & \frac{lc_0}{N \Delta f},  ~k \in \Omega_{l}.
\end{align} 

Define $\mathcal{R}^{\rm N}_{l}$ as the set of search grids for near-field targets in the range cluster $l$ via the 2D-MUSIC algorithm. Each search grid $(d_{\rm pri},\theta_{\rm pri}) \in \mathcal{R}^{\rm N}_{l}$ should also satisfy the above condition (\ref{equ:pri_inf_cluster_l}) by replacing $(x_{k}, y_{k})$ with $(x_{\rm pri}(d_{\rm pri},\theta_{\rm pri}), y_{\rm pri}(d_{\rm pri},\theta_{\rm pri}))$ defined as
\begin{align}
    x_{\rm pri}(d_{\rm pri},\theta_{\rm pri}) &= x_{\rm I} + d_{\rm pri} \cos\theta_{\rm pri}, \\
    y_{\rm pri}(d_{\rm pri},\theta_{\rm pri}) &= y_{\rm I} + d_{\rm pri} \sin\theta_{\rm pri}.
\end{align}
Therefore, the set of search grids for near-field targets in the cluster range $l$ can be presented as
\begin{align}\label{equ:R_N_l}
    \mathcal{R}^{\rm N}_{l} = &~\{ (d,\theta) | (d,\theta) \text{ satisfies conditions (\ref{equ:pri_inf_cluster_l})}, \notag \\
    &~\forall (d,\theta) \in \mathcal{R}^{\rm N} \}, \quad \forall l \in \Phi.
\end{align}

Similarly, for the detection of far-field targets, the number of search grids for the 1D spectrum can also be reduced to save the searching cost with the prior information. The possible positions $(x_{\rm F}, y_{\rm F})$ of far-field targets should satisfy the condition (\ref{equ:pri_inf_cluster_l}) and
\begin{align}
    (d(x_{\rm F}, y_{\rm F}),\theta(x_{\rm F}, y_{\rm F})) \in \mathcal{L}^{\rm F},
\end{align}
where $\mathcal{L}^{\rm F} = \mathcal{L} \backslash \mathcal{L}^{\rm N}$ is the far-field region with $\mathcal{L}$ being the whole position region for targets in the considered system.
Then, from $\mathcal{L}^{\rm F}$, we can obtain the AOA region of the far-field targets in the range cluster $l$ as $\mathcal{A}_l$. Finally, the set of search grids for far-field targets via the MUSIC algorithm can be given by
\begin{align}\label{equ:R_F_l}
    \mathcal{R}^{\rm F}_{l} = \mathcal{A}_l \cap \mathcal{R}^{F}, \quad \forall l \in \Phi.
\end{align}

\begin{figure}[t]
   \centering
    \includegraphics[width=.45\textwidth]{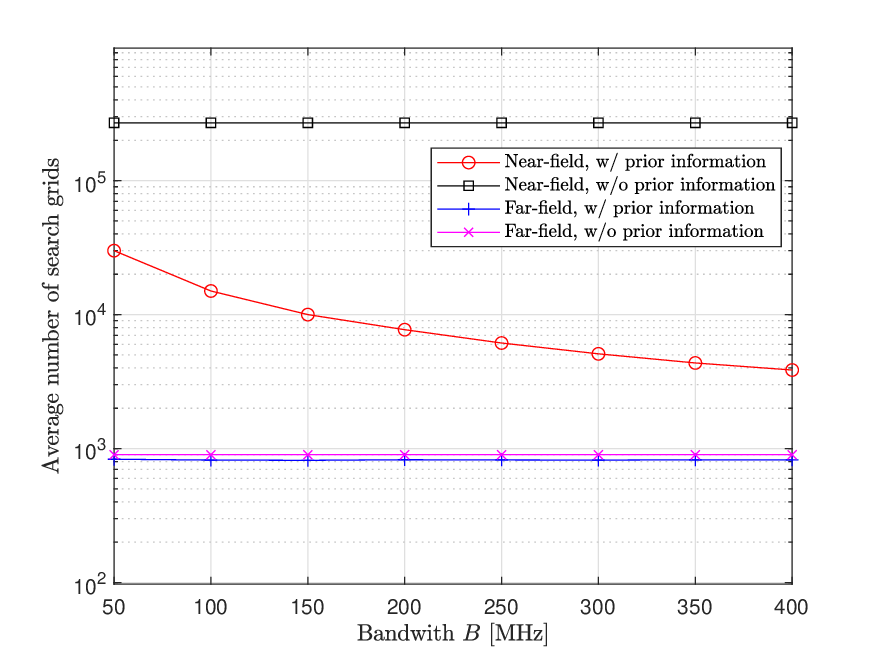}
    \vspace{-0.5cm}
    \caption{Comparison of the average number of search grids for the near-field and far-field targets between the proposed method and the conventional method.}\label{fig:comparison_search_grids}
    \vspace{-0.5cm}
\end{figure}

\begin{remark}
To validate the efficiency by utilizing the prior information, an example is provided in this section. In this example, we assume that there is one user located at $(0,0)$ in meter, one IRS located at $(20,20)$ in meter, and one BS located at $(20,15)$ in meter. Then, we randomly generate the location of one target such that it is located within a quarter circle with the IRS as the center and $d^\text{max}=80$ (meter) as the radius. 
Here, we consider that the near-field region is defined as $\mathcal{L}^{\rm N} = \{(d,\theta)| d \le d_{\rm R} \}$ with $d_{\rm R} = 30$ (meter).
Moreover, we assume the step size of angular search is $0.1$ in degree, i.e., $\Delta \theta = \frac{\pi}{1800}$, and the step size of range search is $0.1$ in meter, i.e., $\Delta d = 0.1$ (meter). As a result, under this setup, the total number of search grids in $\mathcal{R}^{\rm N}$, known as the cardinality of $\mathcal{R}^{\rm N}$, is $|\mathcal{R}^{\rm N}|=900\times 300=2.7 \times 10 ^5$. Last, we adopt the average number of search grids as the performance metric. Fig. \ref{fig:comparison_search_grids} plots the average number of search grids for near-field targets and far-field targets under different bandwidth ranging from $50$ MHz to $400$ MHz. In particular, the utilization of the prior information can significantly reduce the number of search grids for the near-field targets. For example, when the bandwidth is $400$ MHz, the number of the search grids in $\mathcal{R}^{\rm N}$ is $2.7 \times 10^5$ grids on average, while our proposed approach only requires $3858$ grids on average. This result indicates that more than $98$\% search grids can be saved in our proposed method, compared to the conventional approach without considering the prior information. However, for far-field targets, the average number of search grids is only slightly decreased by about $8.8$\% when the bandwidth is $400$ MHz, since the far-field region is much larger than the near-field region. Therefore, utilizing the prior information can mainly save the searching cost for the near-field target detection.
\end{remark}

\begin{algorithm}[t]
	\caption{Prior Information-Assisted MUSIC Algorithm for AOA and Range Estimation in the Range Cluster $l$}\label{alg:pri_music}
	    {\bf Input}: $ \{\boldsymbol{\tilde{h}}_{l,t}^{(q)}\}_{q=1,t=1}^{Q_0,V}$ given in \eqref{eq:estimated_channel cluster l};\\
	    {\bf Initialization (Offline)}: Obtain $\mathcal{R}_l^{\rm N}$'s and $\mathcal{R}_l^{\rm F}$ given in \eqref{equ:R_N_l} and \eqref{equ:R_F_l};
        \begin{enumerate}
        \item [1.] Calculate the sampled covariance matrix $\boldsymbol{\tilde{R}}_l$ via \eqref{equ:sampled_cov_vs_h}, and perform EVD on $\boldsymbol{\tilde{R}}_l$ to obtain eigenvalues $\{\lambda_{l,m}\}_{m=1}^{Q_0M_{\rm B}}$ and the corresponding eigenvectors $\{\boldsymbol{v}_{l,m}\}_{m=1}^{Q_0M_{\rm B}}$; \Comment{Step 1} 
        \item [2.] Estimate the number $\hat{K}_l$ of targets in range cluster $l$ by \eqref{eq:est k}; \Comment{Step 2}
        \item [3.] Search the peaks of the $2$D spectrum \eqref{eq:spectrum 2D} over $\mathcal{R}_{l}^{\rm N}$ and the peaks in $1$D spectrum \eqref{eq:spectrum 1D} over $\mathcal{R}_l^{\rm F}$; \Comment{Step 3}
        \item [4.] Obtain the two sets $\tilde{\mathcal{P}}_{l}^{\rm F}$ and $\tilde{\mathcal{P}}_{l}^{\rm N}$ that contain the largest $\hat{K}_l$ peaks from the $1$D and $2$D spectrums obtained in Step $3$, respectively. Then reserve the peaks larger than the threshold $\varsigma^{\rm F}$ for the set $\tilde{\mathcal{P}}_{l}^{\rm F}$ and reserve the peaks larger than the threshold $\varsigma^{\rm N}$ for the set $\tilde{\mathcal{P}}_{l}^{\rm N}$; \Comment{Step 4}
     \end{enumerate}
        {\bf{Output}}: The AOA information $\hat{\Xi}_l^{\rm F}$ of far-field targets; the AOA and range information $\hat{\Xi}_l^{\rm N}$ of near-field targets.
\end{algorithm}

Until now, we have detected some peaks in $P_l^{\rm F}(\theta)$ and $P_l^{\rm N}(d,\theta)$ for each $l \in \Phi$. However, the number of peaks in $\left\{P_l^{\rm F}(\theta), \forall \theta \in \Xi_l^{\rm F}\right\}$ and $\left\{P_l^{\rm N}(d,\theta), \forall (d,\theta) \in \Xi_l^{\rm N} \right\}$ is usually much larger than $\hat{K}_l$ because some ghost targets may be detected due to noise. To tackle this problem, we propose to determine the final peak sets $\mathcal{P}_l^{\rm F}$ and $\mathcal{P}_l^{\rm N}$ of each range cluster $l \in \Phi$ by thresholding method, which selects all the peaks being larger than the pre-determined thresholds. Note that totally $\hat{K}_l$ targets are considered in each range cluster $l \in \Phi$, meaning that either the number of far-field targets or near-field targets should be no larger than $\hat{K}_l$. Therefore, we first select the largest $\hat{K}_l$ peaks in $P_l^{\rm F}(\theta)$ and the largest $\hat{K}_l$ peaks in $P_l^{\rm N}(d,\theta)$ for each range cluster $l \in \Phi$ to contribute the set $\tilde{\mathcal{P}}^{\rm F}_l$ and $\tilde{\mathcal{P}}^{\rm N}_l$, respectively. Then we utilize two pre-selected thresholds $\varsigma^{\rm F}$ and $\varsigma^{\rm N}$ to determine far-field targets and near-field targets in the range cluster $l \in \Phi$, respectively. Finally, the range and/or AOA information sets for far-field and near-field targets are, respectively, given by
\begin{align}
  \hat{\Xi}_l^{\rm F} &= \{ \theta | P_l^{\rm F}(\theta) > \varsigma^{\rm F}  \text{ and } P_l^{\rm F}(\theta) \in \tilde{\mathcal{P}}_l^{\rm F} \}, \forall l \in \Phi, \label{equ:F_AOA} \\
  \hat{\Xi}_l^{\rm N} &=\{(d,\theta)| P_l^{\rm N}(d,\theta) > \varsigma^{\rm N}  \text{ and } P_l^{\rm N}(d,\theta) \in \tilde{\mathcal{P}}_l^{\rm N} \}, \forall l \in \Phi. \label{equ:N_dAOA}
\end{align}

With the above, the prior information-assisted MUSIC algorithm can be outlined in Algorithm \ref{alg:pri_music}.
After finishing the task in Phase II, target locations will be estimated in Phase III with the estimated AOA and range information of targets.

To summarize, in Phase II, given each range cluster $l \in \Phi$ where exists targets, we 1. estimate the target number in this range cluster as $\hat{K}_l$ given in (\ref{eq:est k}); 2. detect which targets are far-field targets and which targets are near-field targets; 3. estimate the AOAs from the far-field targets to the IRS in the set $\hat{\Xi}_l^{\rm F}$ defined in (\ref{equ:F_AOA}) and the distances and AOAs from the near-field targets to the IRS in the set $\hat{\Xi}_l^{\rm N}$ defined in (\ref{equ:N_dAOA}).

\section{Numerical Results}\label{sec:simulation}

\begin{figure*}[t]
  \centering
  \subfigure[AOA spectrum for far-field targets.]{
    \label{fig:1D_spec}
    \includegraphics[width=.32\textwidth]{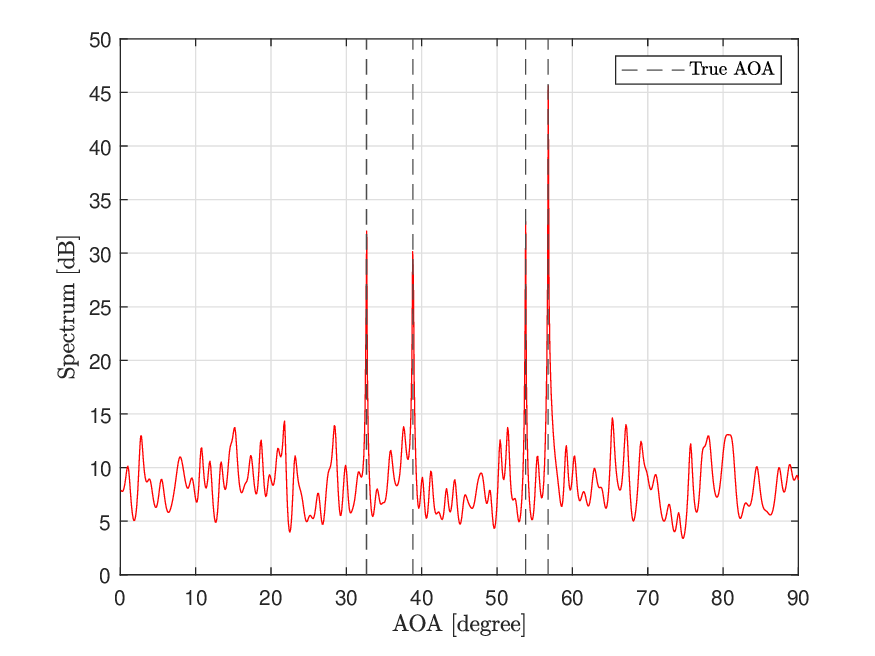}}
  \subfigure[AOA spectrum for near-field targets.]{
    \label{fig:2D_spec_AOA}
    \includegraphics[width=.32\textwidth]{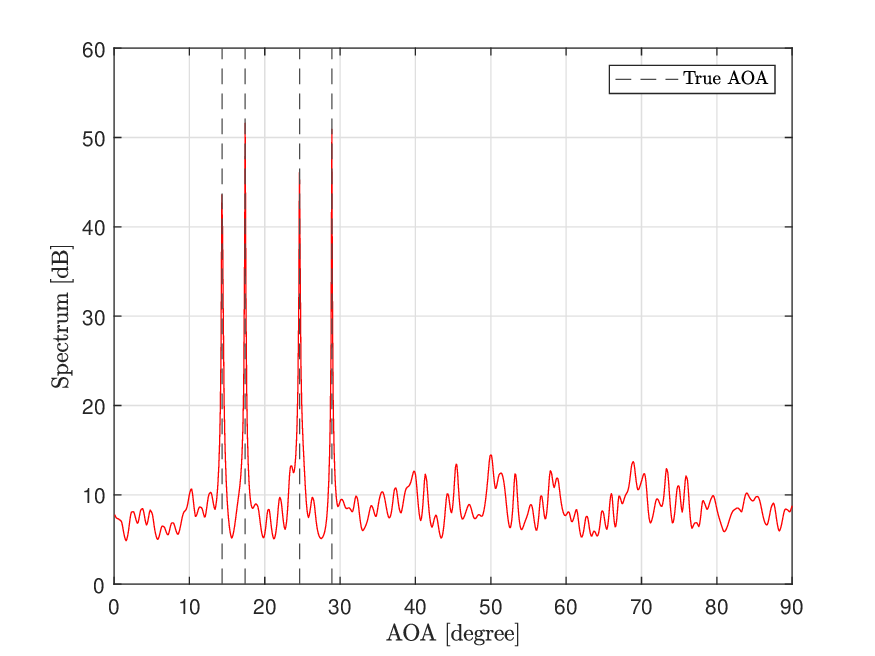}}
  \subfigure[Range spectrum for near-field targets.]{
    \label{fig:2D_spec}
    \includegraphics[width=.32\textwidth]{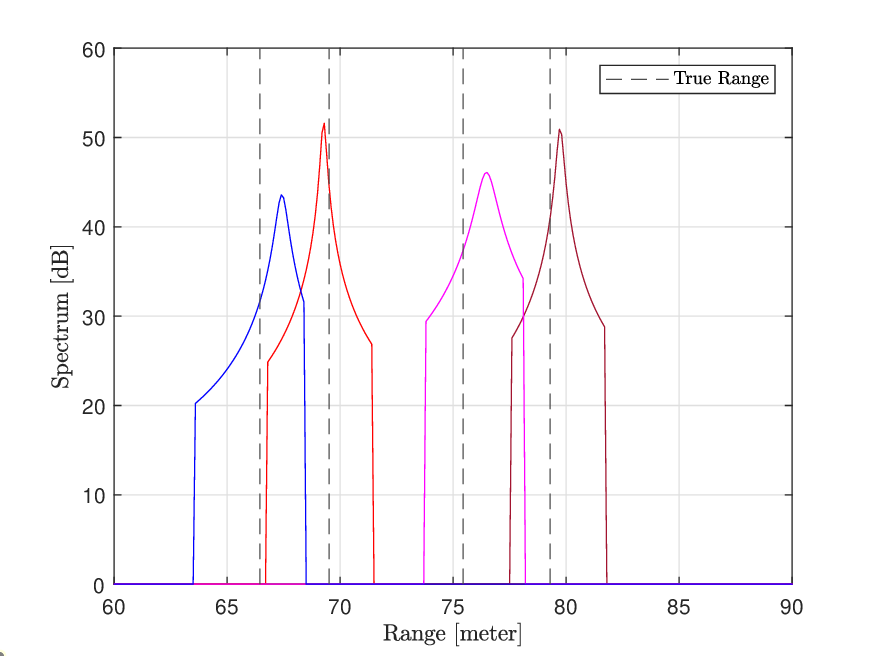}}
    \vspace{-0.3cm}
  \caption{The MUSIC spectrums in Phase II for far-field and near-field targets with $M_{\rm B} = 4$ and $Q_0=4$.}\label{fig:spec}
  \vspace{-0.7cm}
\end{figure*}

In this section, we provide numerical examples to verify the effectiveness of the proposed three-phase localization protocol. The transmit power of the user is $39$ dBm and the power spectrum density of the noise at the receive BS is $-169$ dBm/Hz. We consider $N = 834$ sub-carriers in the OFDM system with bandwidth equaling $100$ MHz and $L=88$ solvable paths. We utilize the received signals in $V=32$ coherence blocks for target localization. Then, the IRS and the BS are modeled as ULA with $\frac{\lambda}{2}$ as the element and antenna spacing. In particular, we assume that the number of the IRS elements is $256$. Next, we assume that the BS, the IRS, and the user are located in $(50,43)$, $(50,50)$, and $(0,0)$ in meter. The near-field region is defined as the circle region centered by the IRS and radius $90$ in meter.
In particular, the challenging case is considered to validate the performance superiority of the proposed method, where there are multiple targets located in the same range cluster.
We generate $\iota_{\rm max}=10^4$ realizations of target locations, while under each realization $\iota$, we consider that targets are located in $3$ range clusters and each of them contains $K=8$ targets randomly either in the near-field or far-field region of the IRS. 

\subsection{Performance Evaluation of Prior Information-Assisted MUSIC in Phase II}

First, we validate the feasibility for extracting the AOA and/or range information by using the prior-information assisted MUSIC algorithm in Phase II in Fig. \ref{fig:spec}. We consider $M_{\rm B}=4$ and $Q_0=4$. Due to the usage of the prior information from the range cluster estimation, the spectrum for the range of the near-field targets only focuses on a narrow region and thus we only need to search the peak in a much small region, which thus reduces the searching cost. It is observed that AOAs of far-field targets can be accurately estimated based on the 1D spectrum, while both of the range and AOA information of near-field targets can be obtained by the 2D spectrum. In addition, it shows that the AOAs of near-field targets can be estimated more accurately than their range information, which reveals that the detection error of near-field targets mainly comes from the range estimation. Since the AOA and/or range information can be extracted via MUSIC, both far-field and near-field targets are able to be localized in Phase III, which also verifies the feasibility to uniquely identify different targets and localize them in the IRS-assisted system.

\begin{figure}
  \centering
  \includegraphics[width=.42\textwidth]{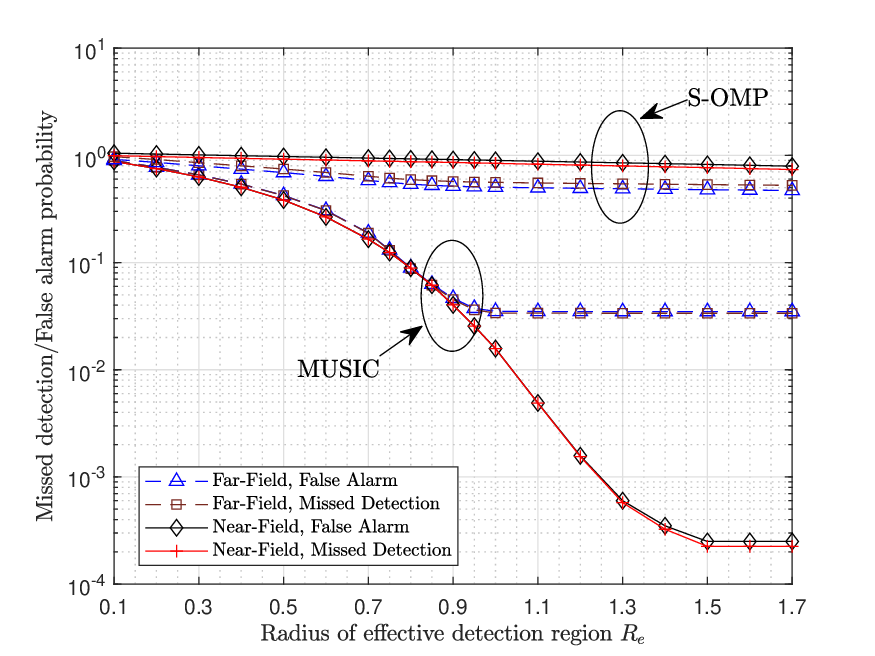}
  \vspace{-0.3cm}
  \caption{Localization performance versus the radius of the effective detection region with $M_{\rm B} = 4$ and $Q_0=4$.}\label{fig:prob_Re}
  \vspace{-0.3cm}
\end{figure}

\subsection{Performance Evaluation of the Three-Phase Protocol}

Two performance metrics will be considered - missed detection and false alarm probabilities for near-field and far-field target detection. 
First, we can define the effective detection region for each true target $k$ as a circle region centered by itself with radius $R_{e}$. Specifically, under the each realization, we claim that a missed detection event occurs for a near-field (far-field) target $k$ if the locations of all the detected near-field (far-field) targets estimated by our proposed method lie outside the effective detection region of this target $k$. Moreover, we claim that a false alarm event occurs for a detected near-field (far-field) target $k$ if its estimated location lies outside the effective detection region of all the near-field (far-field) targets. 
Let $K_{\iota}^{\rm N}$ and $K_{\iota}^{\rm F}$ denote the numbers of near-field and far-field targets generated in realization $\iota$, $K_{\iota}^{\rm N,MD}$, $K_{\iota}^{\rm F,MD}$, $K_{\iota}^{\rm N,FA}$, and $K_{\iota}^{\rm F,FA}$ denote the number of missed detection events for near-field targets and those for far-field targets, and false alarm events for near-field targets and those for far-field targets at realization $\iota$. Then, after $\iota_{\rm max} = 10^4$ realizations, the missed detection probabilities for detecting the near-field and far-field targets are calculated as $P_{\text{MD}}^{\rm N} = \frac{\sum_{\iota=1}^{\iota_\text{max}}K^{{\rm N,MD}}_{\iota}}{\sum_{\iota=1}^{\iota_{\text{max}}} K_{\iota}^{\rm N}}$ and $P_{\text{MD}}^{\rm F} = \frac{\sum_{\iota=1}^{\iota_\text{max}}K^{{\rm F,MD}}_{\iota}}{\sum_{\iota=1}^{\iota_{\text{max}}} K_{\iota}^{\rm F}}$, and the false alarm probabilities for detecting the near-field and far-field targets are defined as $P^{\rm N}_{\text{FA}} = \frac{\sum_{\iota=1}^{\iota_\text{max}}K^{{\rm N,FA}}_{\iota}}{\sum_{\iota=1}^{\iota_{\text{max}}} K_{\iota}^{\rm N}}$ and $P^{\rm F}_{\text{FA}} = \frac{\sum_{\iota=1}^{\iota_\text{max}}K^{{\rm F,FA}}_{\iota}}{\sum_{\iota=1}^{\iota_{\text{max}}} K_{\iota}^{\rm F}}$. The representative sparsity-based algorithm of simultaneous orthogonal matching pursuit (S-OMP) \cite{Tropp_2006_SP} is also served as the benchmark scheme. 

\begin{figure}[t]
  \centering
  \includegraphics[width=.42\textwidth]{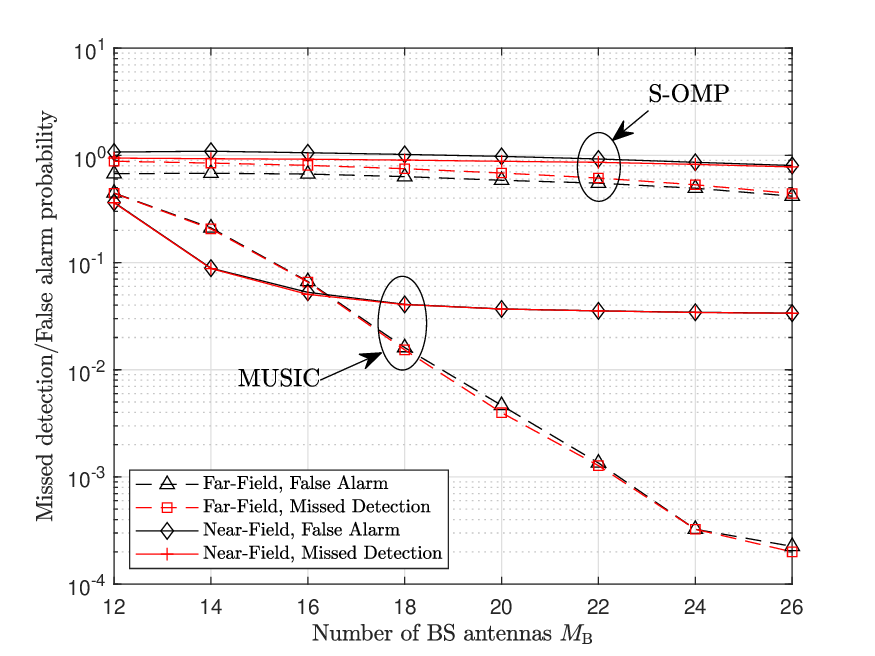}
  \vspace{-0.3cm}
  \caption{Localization performance versus the number of BS antennas $M_{\rm B}$ with $Q_0=1$ when the BS-IRS channel follows near-field channel model.}\label{fig:prob_MB}
  \vspace{-0.3cm}
\end{figure}

\begin{figure}[t]
  \centering
  \includegraphics[width=.42\textwidth]{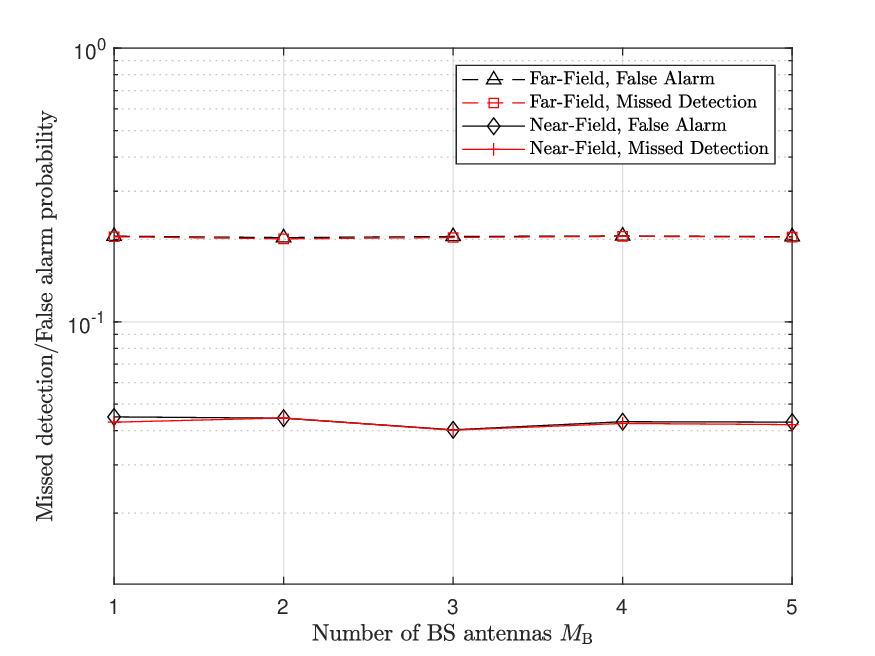}
  \vspace{-0.3cm}
  \caption{Localization performance versus the number of BS antennas $M_{\rm B}$ with $Q_0=10$ when the BS-IRS channel follows far-field channel model.}\label{fig:prob_MB_far-BS-IRS}\vspace{-0.5cm}
\end{figure}

The impact of the detection radius $R_{e}$ on the performance\footnote{The selection of thresholds $\varsigma^{\rm F}$ and $\varsigma^{\rm N}$ realizes a trade-off between the performance of false alarm probability and missed detection probability. In simulations, thresholds $\varsigma^{\rm F}$ and $\varsigma^{\rm N}$ are chosen to realize close performance of false alarm probability and missed detection probability for fair comparison.} of the proposed localization protocol is first evaluated in Fig. \ref{fig:prob_Re} with $M_{\rm B} = 4$ and $Q_0=4$. We can observe that enlarging the detection region $R_e$ can enhance the localization performance of both near-field and far-field targets for both the MUSIC and S-OMP algorithms. This improvement, however, comes with an increases in the tolerable positioning error. Additionally, the MUSIC algorithm can significantly outperform the S-OMP algorithm, since inadequate number of measurements usually leads to limited performance for CS-based algorithms. It is also shown that the missed detection/false alarm probability of far-field and near-field targets approaches their minimal value for $R_e \ge 1.1$ (meter) and $R_e \ge 1.5 $ (meter), respectively. If we set $R_e = 1$ (meter), missed detection and false alarm probabilities of far-field and near-field targets are all less than 0.04, indicating that precise meter-level localization performance can be realized via the proposed three-phase protocol. In the following simulations, we set $R_e = 1$ (meter).

\begin{figure}[t]
  \centering
  \includegraphics[width=.42\textwidth]{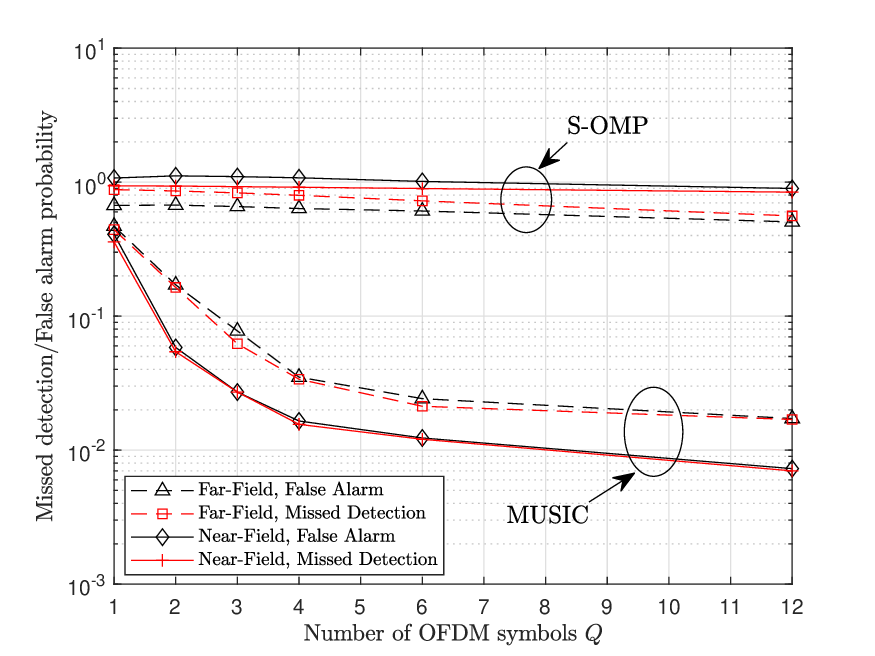}
  \vspace{-0.3cm}
  \caption{Localization performance versus the number of utilized OFDM symbols $Q_0$ used in each coherence block for localization under $Q_0 M_{\rm B} = 12$.}\label{fig:prob_Q}
  \vspace{-0.3cm}
\end{figure}

\begin{figure}[t]
  \centering
  \includegraphics[width=.42\textwidth]{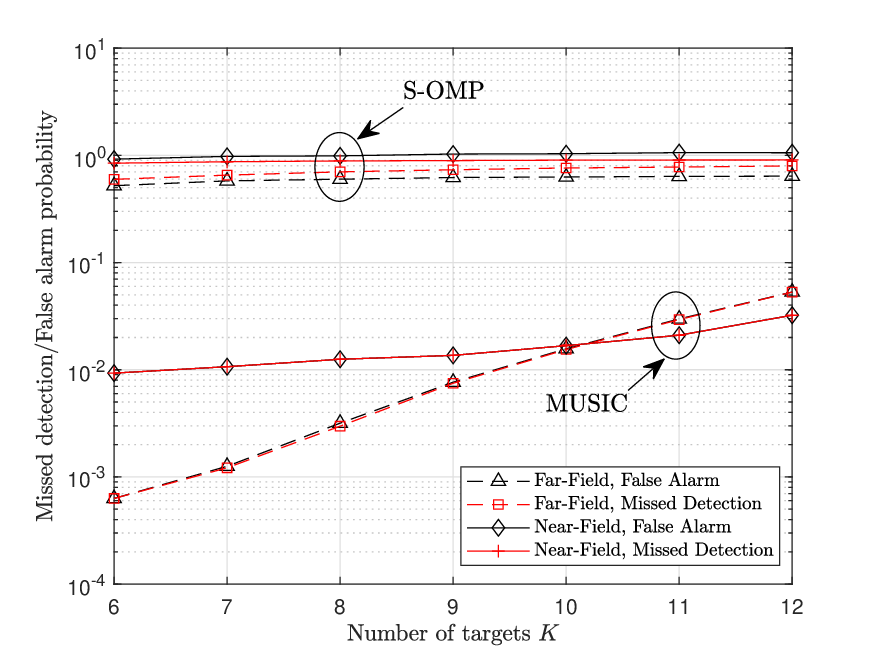}
  \vspace{-0.3cm}
  \caption{Localization performance versus the number of targets $K$ with $M_{\rm B} = 4$ and $Q_0=4$.}\label{fig:prob_K}\vspace{-0.5cm}
\end{figure}

We then evaluate the localization performance under different numbers of BS antennas $M_{\rm B}$ in Fig. \ref{fig:prob_MB} with $Q_0=1$ when the BS-IRS channel follows the near-field channel model. It is observed that both of false alarm and missed detection probabilities for near-field and far-field targets are reduced with the $M_{\rm B}$ increases, because more measurements for target localization are obtained. 
In particular, the missed detection and false alarm probabilities of far-field targets has much more reduction with $M_{\rm B}$ increasing, compared with near-field targets. This is because increasing $M_{\rm B}$ can provide significant performance improvement on the AOA estimation while the range information estimation performance is only slightly improved, which is similar to results in Fig. \ref{fig:spec}. We also consider the case where the BS-IRS channel follows the far-field channel model in Fig. \ref{fig:prob_MB_far-BS-IRS}. It is observed that solely increasing the number of BS antennas cannot provide any enhancement in localization accuracy, which validates our previous analysis that increasing the number of BS antennas in such case only provides redundant information for localization.

The impact of the number of utilized OFDM symbols $Q_0$ in each coherence block on the localization performance is shown in Fig. \ref{fig:prob_Q}. Here, we fix the dimension of measurements with $Q_0 M_{\rm B} = 12$. It is observed that increasing the number $Q_0$ of utilized OFDM symbols used in each coherence block can significantly boost the localization performance. This is because the target detection performance is limited to the ill-conditioned matrix $\breve{\boldsymbol{\Psi}}(\Theta_l)$ resulting from the channel matrix $\boldsymbol{G}$ when $Q_0$ is small. By increasing $Q_0$, the matrix $\breve{\boldsymbol{\Psi}}(\Theta_l)$ can become a better-conditioned matrix and thus the target detection performance is enhanced. These results also indicate that we can use more OFDM symbols in each coherence block to reduce the localization error if the number of BS antennas is small due to hardware limits.

We also evaluate the localization performance of the proposed protocol under different target numbers in Fig. \ref{fig:prob_K} with $M_{\rm B} = 4$ and $Q_0=4$. It is observed that missed detection and false alarm probabilities are both enlarged with the number $K$ of targets increases. With more targets existing in the system, more reflection signals by targets are interfered with each other, leading to degraded performance. Therefore, we should utilize more OFDM symbols in each coherence block to maintain the performance when there are target number is very large in the system, as revealed in Fig. \ref{fig:prob_Q}.

\section{Conclusions}

In this paper, we considered the localization problem for passive targets in an IRS-assisted bi-static 6G ISAC network, where LOS paths between the targets and the BS do not exist, and the IRS serves as a passive anchor to localize the targets. The main challenges lied in how to apply the subspace-based method to jointly localize these mixed far-field and near-field targets from the received signals at the BS. To tackle these challenges, we proposed a three-phase localization protocol that is able to exploit the BS received signals to detect which targets are in the near-field region and which are in the far-field region of the IRS, and localize these targets. Numerical results were provided to verify the effectiveness of our proposed three-phase localization protocol, which demonstrate the feasibility of exploiting the IRSs as passive anchors for the localization of passive targets in the future 6G network to realize ISAC.

\begin{appendices}
\section{Proof of Theorem \ref{theorem1}}\label{appendix1}

First, it is easy to know that the matrix $\boldsymbol{P}$ should satisfy $\text{rank}(\boldsymbol{P}) \ge K^{\rm max}$ due to the inequality $\boldsymbol{\breve{\Psi}}(\Theta_l) \le \min\{\boldsymbol{P}, \boldsymbol{A}_{\rm I}(\Theta_l)\}$. Since $\text{rank}(\boldsymbol{P}) \le Q_0 r_{\boldsymbol{G}}$ and $r_{\boldsymbol{G}} \ge 1$, condition 1) holds to ensure that $\text{rank}(\boldsymbol{P}) \ge K^{\rm max}$ can be realized in both the cases where the BS-IRS channel follows the far-field and near-field channel models. Then, the condition 2) is to ensure that $K^{\rm max}$ distinct targets can be identified without ambiguity since any $M_{\rm I}$ steering vectors in ULA array manifold are linearly independent \cite{linear_indpend}. The condition 2) also supports that the condition 3) can be realized. Finally, assuming that conditions 1) and 2) both hold, based on the reflecting pattern design in \eqref{equ:IRS_phi_design}, the first row of each matrix $\boldsymbol{P}^{(q)}$ can be expressed as $\boldsymbol{p}^{(q)}_{1,:} = (\boldsymbol{w}_{q}^{\rm I})^T\boldsymbol{D}(\vartheta)$ with $\boldsymbol{D}(\vartheta) = \text{diag}([1,\dots,e^{j(M_{\rm I}-1)\vartheta}])$. Define the matrix $\boldsymbol{P}_1 = [(\boldsymbol{p}^{(1)}_{1,:})^T,\dots,(\boldsymbol{p}^{(Q_0)}_{1,:})^T]^T$. We can get the relationship
\begin{align}\label{equ:col_space_P1}
    \mathcal{N}(\boldsymbol{P}_1) = \text{span}(\boldsymbol{D}(\vartheta)\boldsymbol{w}_{Q_0 + 1}^{\rm I},\dots,\boldsymbol{D}(\vartheta)\boldsymbol{w}_{M_{\rm I}}^{\rm I}).
\end{align}
We can then follow \cite[Proposition 1]{Amini_2005_SPL} to prove that the matrix $\boldsymbol{\breve{\Psi}}_1(\Theta_l) = \boldsymbol{P}_1 \boldsymbol{A}_{\rm I}(\Theta_l)$ is full-column rank with $\text{rank}(\boldsymbol{\breve{\Psi}}_1(\Theta_l)) = K_l$ by contraction if $\boldsymbol{D}(\vartheta) \boldsymbol{w}_{m}^{\rm I} \ne \boldsymbol{a}_{\rm I}(\bar{d}_k,\theta_k))$, $\forall m \in \{Q_0+1,\dots,M_{\rm I}\}, k \in \{1,\dots,K\}$.
However, it is easy to know that the probability 
\begin{align}
    &\text{Pr}(\boldsymbol{D}(\vartheta) \boldsymbol{w}_{m}^{\rm I} = \boldsymbol{a}_{\rm I}(\bar{d}_k,\theta_k)) = 0, \notag \\
    &\qquad\qquad \forall m \in \{Q_0+1,\dots,M_{\rm I}\}, k \in \{1,\dots,K\}.
\end{align}
Due to the fact $\mathcal{N}(\boldsymbol{P}) \subseteq \mathcal{N}(\boldsymbol{P}_1)$, we can obtain that $\text{rank}(\boldsymbol{\breve{\Psi}}(\Theta_l)) = K_l$ also holds almost surely. 
Therefore, the probability of $\text{rank}(\boldsymbol{\breve{\Psi}}(\Theta_l)) = K_l$ is equal to one.
The proof is completed.

\end{appendices}

\bibliographystyle{IEEEtran}
\bibliography{ref}

\end{document}